\documentstyle[12pt,twoside,epsfig,fleqn,espcrc1]{article}

\newcommand{\AmS}{{\protect\the\textfont2
  A\kern-.1667em\lower.5ex\hbox{M}\kern-.125emS}}
\begin{document}
\def\ap#1#2#3   {{\em Ann. Phys. (NY)} {\bf#1} (#2) #3}   
\def\apj#1#2#3  {{\em Astrophys. J.} {\bf#1} (#2) #3.} 
\def\apjl#1#2#3 {{\em Astrophys. J. Lett.} {\bf#1} (#2) #3.}
\def\app#1#2#3  {{\em Acta. Phys. Pol.} {\bf#1} (#2) #3.}
\def\ar#1#2#3   {{\em Ann. Rev. Nucl. Part. Sci.} {\bf#1} (#2) #3.}
\def\cpc#1#2#3  {{\em Computer Phys. Comm.} {\bf#1} (#2) #3.}
\def\epj#1#2#3  {{\em Europ. Phys. J.} {\bf#1} (#2) #3}
\def\err#1#2#3  {{\it Erratum} {\bf#1} (#2) #3.}
\def\ib#1#2#3   {{\it ibid.} {\bf#1} (#2) #3.}
\def\jmp#1#2#3  {{\em J. Math. Phys.} {\bf#1} (#2) #3.}
\def\ijmp#1#2#3 {{\em Int. J. Mod. Phys.} {\bf#1} (#2) #3}
\def\jetp#1#2#3 {{\em JETP Lett.} {\bf#1} (#2) #3}
\def\jpg#1#2#3  {{\em J. Phys. G.} {\bf#1} (#2) #3.}
\def\mpl#1#2#3  {{\em Mod. Phys. Lett.} {\bf#1} (#2) #3.}
\def\nat#1#2#3  {{\em Nature (London)} {\bf#1} (#2) #3.}
\def\nc#1#2#3   {{\em Nuovo Cim.} {\bf#1} (#2) #3.}
\def\nim#1#2#3  {{\em Nucl. Instr. Meth.} {\bf#1} (#2) #3.}
\def\np#1#2#3   {{\em Nucl. Phys.} {\bf#1} (#2) #3}
\def\pcps#1#2#3 {{\em Proc. Cam. Phil. Soc.} {\bf#1} (#2) #3.}
\def\pl#1#2#3   {{\em Phys. Lett.} {\bf#1} (#2) #3}
\def\prep#1#2#3 {{\em Phys. Rep.} {\bf#1} (#2) #3}
\def\prev#1#2#3 {{\em Phys. Rev.} {\bf#1} (#2) #3}
\def\prl#1#2#3  {{\em Phys. Rev. Lett.} {\bf#1} (#2) #3}
\def\prs#1#2#3  {{\em Proc. Roy. Soc.} {\bf#1} (#2) #3.}
\def\ptp#1#2#3  {{\em Prog. Th. Phys.} {\bf#1} (#2) #3.}
\def\ps#1#2#3   {{\em Physica Scripta} {\bf#1} (#2) #3.}
\def\rmp#1#2#3  {{\em Rev. Mod. Phys.} {\bf#1} (#2) #3}
\def\rpp#1#2#3  {{\em Rep. Prog. Phys.} {\bf#1} (#2) #3.}
\def\sjnp#1#2#3 {{\em Sov. J. Nucl. Phys.} {\bf#1} (#2) #3}
\def\shep#1#2#3 {{\em Surveys in High Energy Phys.} {\bf#1} (#2) #3}
\def\spj#1#2#3  {{\em Sov. Phys. JEPT} {\bf#1} (#2) #3}
\def\spu#1#2#3  {{\em Sov. Phys.-Usp.} {\bf#1} (#2) #3.}
\def\zp#1#2#3   {{\em Zeit. Phys.} {\bf#1} (#2) #3}

\hyphenation{author another created financial paper re-commend-ed}

\title{The proton and the photon, who is probing whom in electroproduction?}
\author{Aharon Levy\thanks{This work was partially supported by the 
German--Israel Foundation(GIF), by the U.S.--Israel Binational
Foundation (BSF) and by the Israel Science Foundation (ISF). The
financial support of my visit to Japan by JSPS is highly appreciated.}
\\ \vspace*{0.3cm}
School of Physics and Astronomy\\Raymond and Beverly Sackler Faculty
of Exact Sciences\\ Tel Aviv University, Tel Aviv, Israel}

\maketitle

\begin{abstract}
The latest results on the structure of the proton and the photon as
seen at HERA are reviewed while discussing the question posed in the
title of the talk.
\end{abstract}

\section{INTRODUCTION}
The HERA collider, where 27.5 GeV electrons collide with 920 GeV
protons, is considered a natural extension of Rutherford's experiment
and the process of deep inelastic $ep$ scattering (DIS) is interpreted
as a reaction in which a virtual photon, radiated by the incoming
electron, probes the structure of the proton. In this talk I would
like to discuss this interpretation and ask the question of who is
probing whom~\cite{who}.

The structure of the talk will be the following: it will start with
posing the problem, after which our knowledge about the structure of
the proton as seen at HERA~\cite{rmp} will be presented followed by a
description of our present understanding of the structure of the
photon as seen at HERA and at LEP~\cite{jon,nisius}. Next, an answer
to the question posed in the title will be suggested and the talk will
be concluded by some remarks about the nature of the interaction
between the virtual photon and the proton~\cite{troika}.
\section{THE QUESTION - WHO IS PROBING WHOM?}
\subsection{The process of DIS}
\begin{figure}
\begin{minipage}{7.5cm}
\epsfig{file=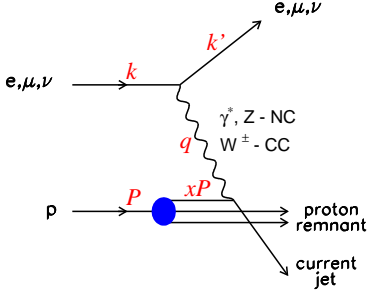,height=0.25\vsize}
\end{minipage}
\begin{minipage}{7.5cm}
\vspace*{-1.5cm}
\caption{A diagram describing the process of deep inelastic scattering (DIS). 
The four vector of the incoming and outgoing leptons are $k$ and
$k^\prime$, that of the exchanged boson is $q$, and that of the
incoming proton is $P$. The four momentum of the struck quark is $xP$.
}
\label{fig:kinematics}
\end{minipage}
\end{figure}

The process of DIS is usually represented by the diagram shown in
figure~\ref{fig:kinematics}. 
If the lepton does not change its
identity during the scattering process, the reaction is labeled
neutral current (NC), as either a virtual photon or a $Z^0$ boson can
be exchanged. When the identity of the lepton changes in the process,
the reaction is called charged current (CC) and a charged $W^\pm$
boson is exchanged. During this talk we will discuss only NC processes. 
Using the four vectors as indicated in the figure, one can define the
usual DIS variables: $Q^2=-q^2$, the 'virtuality' of the exchanged
boson, $x=Q^2/(2P\cdot q)$, the fraction of the proton momentum
carried by the interacting parton, $y=(P\cdot q)/(P\cdot k)$, the
inelasticity, and $W^2=(q+P)^2$, the boson-proton center of mass
energy squared.

The interpretation of the diagram describing a NC event is the
following. The electron beam is a source of photons with virtuality
$Q^2$. These virtual photons `look' at the proton. Any `observed'
structure belongs to the proton. How can we be sure that we are indeed
measuring the structure of the proton? Virtual photons have no
structure. Is that always true? We know that real photons have
structure; we even measure the photon structure function
$F_2^\gamma$~\cite{nisius}. Let us discuss this point further in the
next subsections.
\subsection{The fluctuating photon}
\begin{figure}[htb]
\vspace*{-1cm}
\begin{minipage}{7.5cm}
\epsfig{file=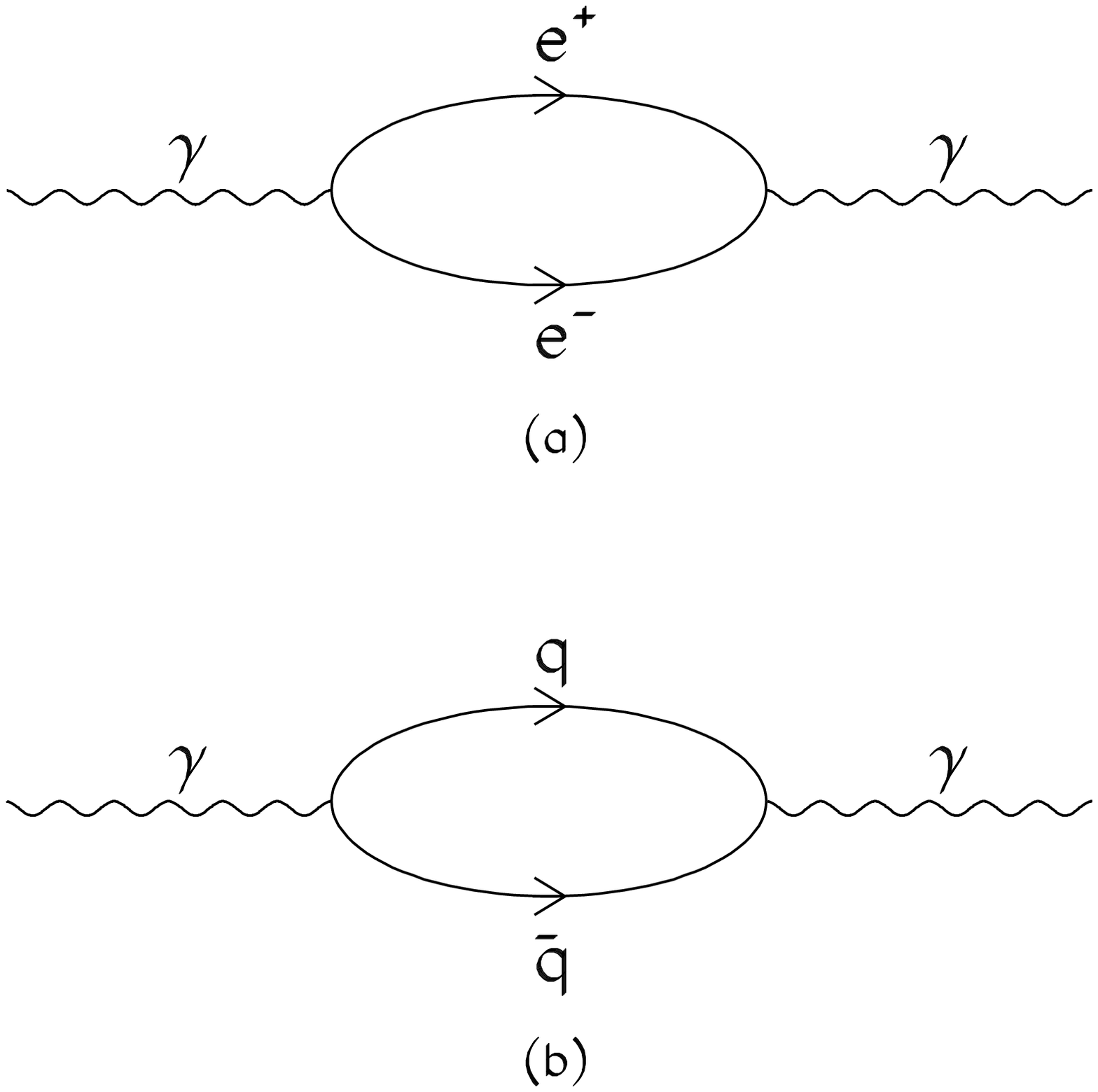,height=0.25\vsize}
\vspace*{-1.1cm}
\caption{Fluctuation of a photon into (a) an $e^+e^-$ pair, (b) a $q\bar{q}$ 
pair.}
\label{fig:fluc}
\end{minipage}
\hspace*{2mm}
\vspace*{-1cm}
\begin{minipage}{7.5cm}
\epsfig{file=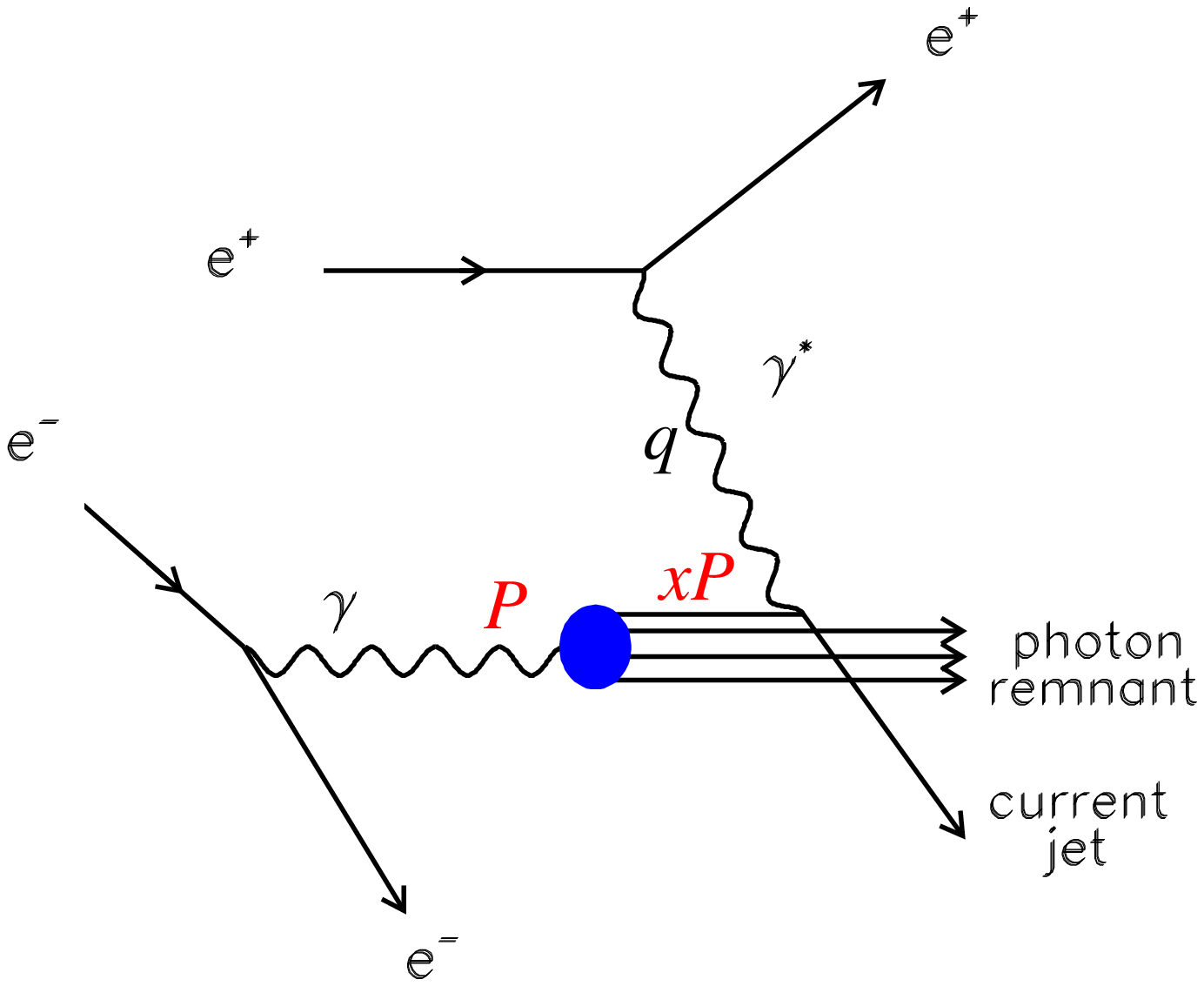,height=0.25\vsize}
\vspace*{-1.2cm}
\caption{A diagram describing a DIS process on a quasi-real photon using the 
reaction $e^+e^- \to e^+e^-X$.}
\label{fig:dis-photon}
\end{minipage}
\end{figure}

How is it possible that the photon, which is the gauge particle
mediating the electromagnetic interactions, has a hadronic structure?
Ioffe's argument~\cite{ioffe}: the photon can fluctuate into
$q\bar{q}$ pairs just like it fluctuates into $e^+e^-$ pairs (see
figure~\ref{fig:fluc}).  If the fluctuation time, defined in the
proton rest frame as $t_f \simeq (2E_\gamma)/m^2_{q\bar{q}}$, is much
larger than the interaction time, $t_{int} \simeq r_p$, the photon
builds up structure in the interaction. Here, $E_\gamma$ is the energy
of the fluctuating photon, $m_{q\bar{q}}$ is the mass into which it
fluctuates, and $r_p$ is the radius of the proton.
The hadronic structure of the photon, built during the interaction,
can be studied by measuring the photon structure function $F_2^\gamma$
in a DIS type of experiment where a quasi-real photon is probed by a
virtual photon, both of which are emitted in $e^+e^-$ collisions, as
described in figure~\ref{fig:dis-photon}.  This diagram is very
similar to that in DIS on a proton target
(figure~\ref{fig:kinematics}).
\subsection{Structure of virtual photons?}
Does a virtual photon also fluctuate and acquire a hadronic structure?
The fluctuation time of a photon with virtuality $Q^2$ is given by
$t_f \simeq (2E_\gamma)/(m^2_{q\bar{q}}+Q^2)$, and thus at very high
$Q^2$ one does not expect the condition $t_f \gg t_{int}$ to
hold. However at very large photon energies, or at very low $x$, the
fluctuation time is independent of $Q^2$: $t_f \simeq 1/(2m_px)$,
where $m_p$ is the proton mass, and thus even highly virtual photons
can acquire structure. For instance, at HERA presently $W \sim$ 200 -
300 GeV, and since $x \approx Q^2/(Q^2+W^2)$, $x$ can be as low as
0.01 even for $Q^2$ = 1000 GeV$^2$. In this case, the fluctuation time
will be very large compared to the interaction time and the highly
virtual photon will acquire a hadronic structure. How do we interprate
the DIS diagram of figure~\ref{fig:kinematics} in this case? Whose
structure do we measure? Do we measure the structure of the proton,
from the viewpoint of the proton infinite momentum frame, or do we
measure the structure of the virtual photon, from the proton rest
frame view?  Who is probing whom?

When asked this question, Bjorken answered~\cite{bj} that physics
can not be frame dependent and therefore it doesn't matter: we can say
that we measure the structure of the proton or we can say that we
study the structure of the virtual photon. I will try to convince you
at the end of my talk that this answer makes sense.
\section{THE STRUCTURE OF THE PROTON}
In this section we will refrain from discussing the question posed
above and will accept the interpretation of measuring the structure of
the proton via the DIS diagram in 
figure~\ref{fig:kinematics}. 
We
present below information about the structure of the proton as seen
from the DIS studies at HERA.
\subsection{HERA}
\begin{figure}[htb]
\begin{minipage}{8.5cm}
\epsfig{file=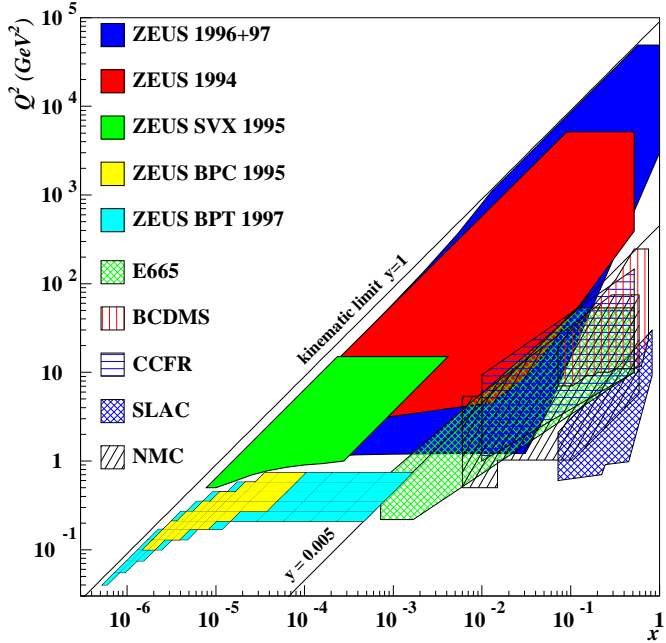,height=8.5cm}
\end{minipage}
\hspace{3mm}
\begin{minipage}{6.5cm}
\vspace*{-1.1cm}
\caption{The $x$-$Q^2$ kinematic plane of some of the fixed target and of 
the HERA collider DIS experiments.}
\label{fig:bpt}
\end{minipage}
\end{figure}

With the advent of the HERA $ep$ collider the kinematic plane of
$x$-$Q^2$ has been extended by 2 orders of magnitude in both variables
from the existing fixed target DIS experiments, as depicted in
figure~\ref{fig:bpt}.

The DIS cross section for $ep \to eX$ can be written (for $Q^2 \ll
M_Z^2$) as~\cite{leader-predazzi},
\begin{equation}
\frac{d^2\sigma}{dxdQ^2}=\frac{4\pi\alpha^2}{xQ^2}\left\{\frac{y^2}{2}
2xF_1(x,Q^2)+(1-y)F_2(x,Q^2)\right\}.
\end{equation}
In the quark-parton model (QPM), the proton structure function $F_2$
is only a function of $x$ and can be expressed as a sum of parton
densities, and is related to $F_1$ through the Callan-Gross
relation~\cite{cg},
\begin{equation}
F_2(x)=\sum_ie_i^2xq_i(x)=2xF_1,
\end{equation}
where $e_i$ is the electric charge of quark $i$ and the index $i$ runs
over all the quark flavours.

Note that in Quantum Chromodynamics (QCD), the Callan-Gross relation
is violated, and the structure function is a function of $x$ and
$Q^2$,
\begin{equation}
F_2(x,Q^2)-2xF_1(x,Q^2)=F_L(x,Q^2)>0,
\end{equation}
where the longitudinal structure function $F_L$ contributes in an
important way only at large $y$.

The motivation for measuring $F_2(x,Q^2)$ can be summarized as
follows: (a) test the validity of perturbative QCD (pQCD)
calculations, (b) decompose the proton into quarks and gluons, and (c)
search for proton substructure.
\subsection{QCD evolution - scaling violation}
\begin{figure}[htb]
\begin{minipage}{7.5cm}
\epsfig{file=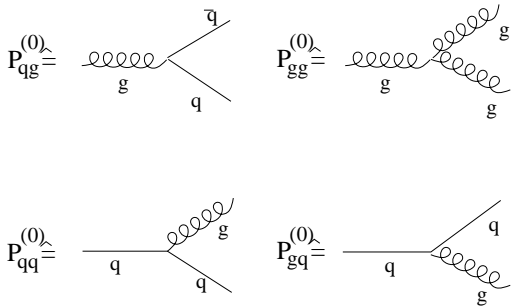,height=4cm}
\end{minipage}
\hspace*{2mm}
\begin{minipage}{7.5cm}
\vspace*{-1cm}
\caption{Splitting functions $P_{ij}$ in leading order, describing the 
splitting of parton $j$ into parton $i$.}
\label{fig:splitting}
\end{minipage}
\end{figure}

\begin{figure}[htb]
\vspace{-1cm}
\begin{center}
\epsfig{file=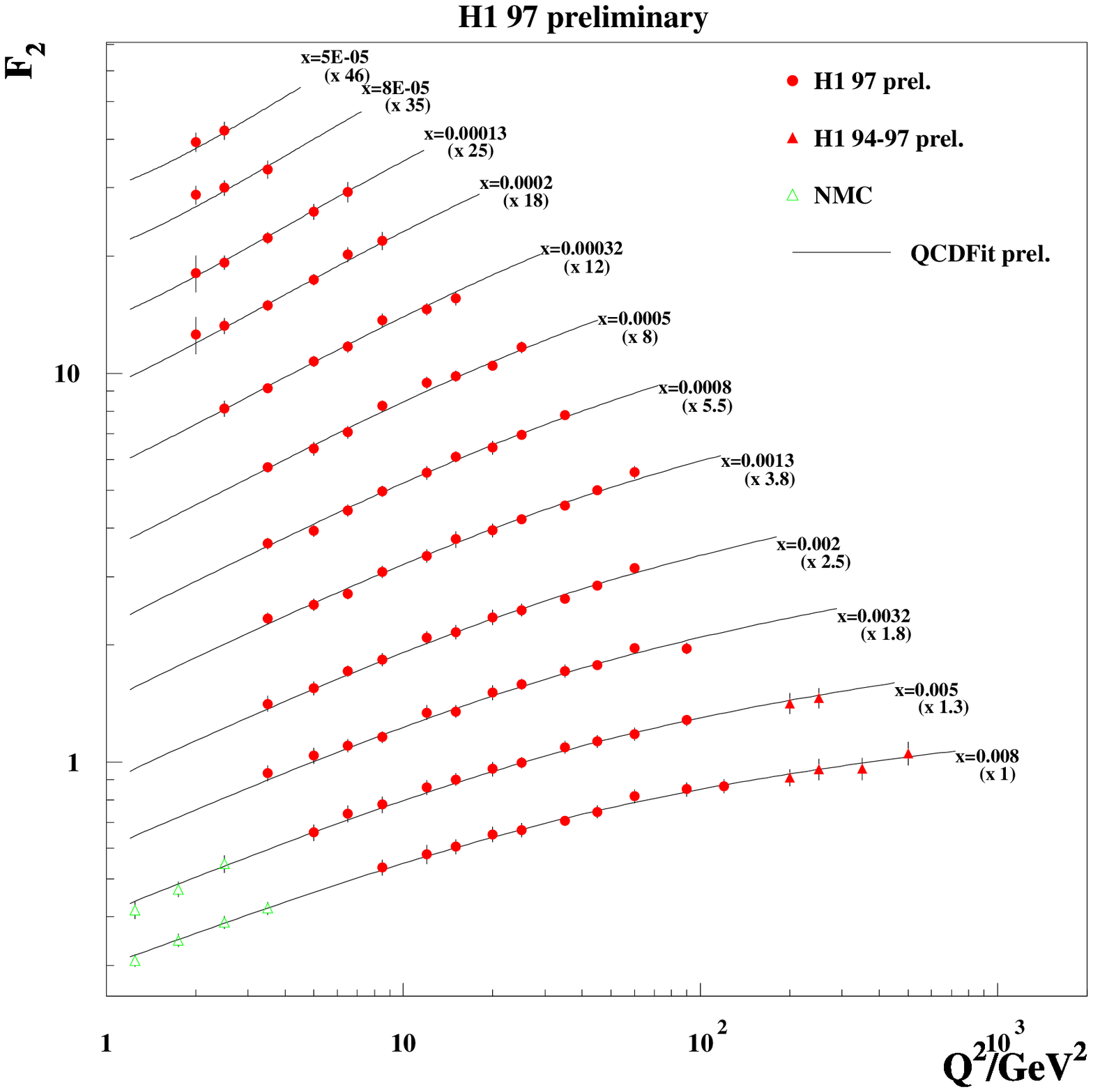,height=6.5cm}
\epsfig{file=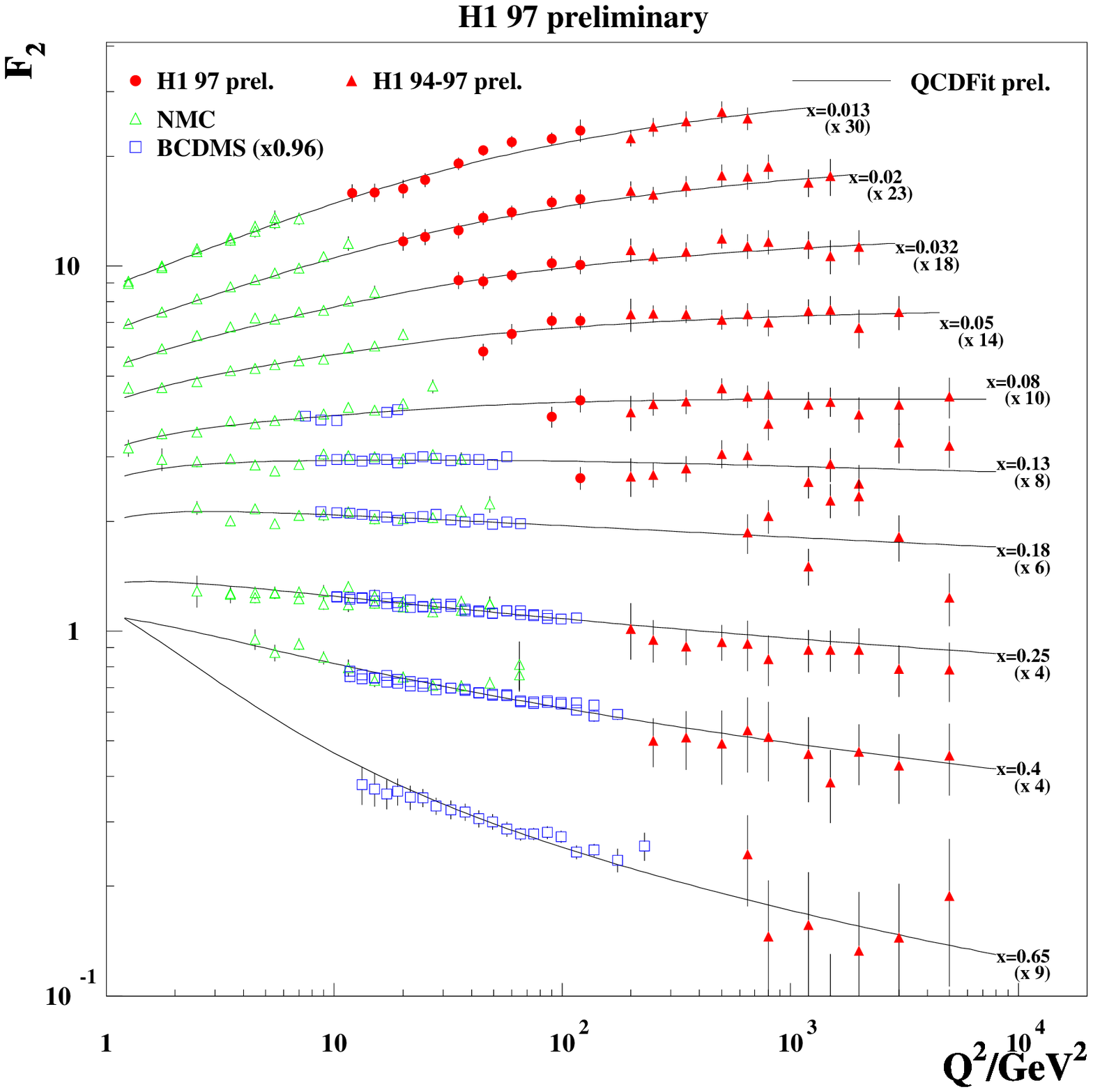,height=6.5cm}
\end{center}
\vspace*{-1.1cm}
\caption{Comparison of the scaling violation behaviour of $F_2$ with the 
results of a next-to-leading order DGLAP evolution equation.}
\label{fig:sc-viol}
\end{figure}

Quarks radiate gluons; gluons split and produce more gluons at low $x$
and also $q\bar{q}$ pairs at low $x$. This QCD evolution chain is
usually described in leading order by splitting functions $P_{ij}$, as
shown in 
figure~\ref{fig:splitting}.
This procedure leads to scaling violation in the following way: there
is an increase of $F_2$ with $Q^2$ at low $x$ and a decrease at high
$x$. Scaling holds at about $x$=0.1. The data follows this prediction
of QCD, as can be seen in 
figure~\ref{fig:sc-viol}.
\subsection{Overview of $F_2$}

\begin{figure}[htb]
\begin{minipage}{7.5cm}
\epsfig{file=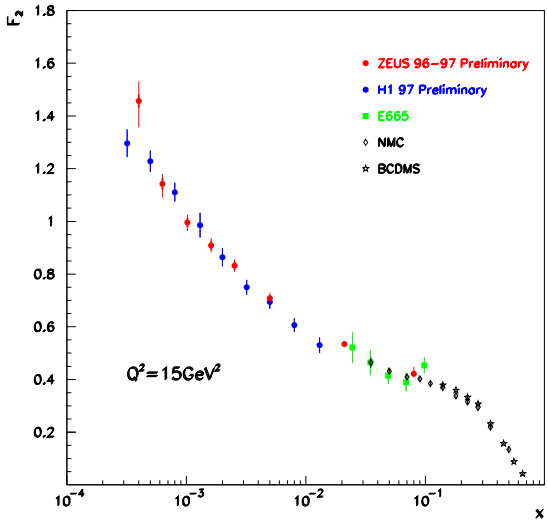,height=7cm}
\end{minipage}
\hspace*{5mm}
\begin{minipage}{7.5cm}
\vspace*{-1cm}
\caption{The proton structure function $F_2$, as function of $x$, at 
$Q^2$ = 15 GeV$^2$, for HERA and some fixed target data.}
\label{fig:f2-over}
\end{minipage}
\end{figure}

The fixed target experiments provided information at relatively high
$x$ and thus enabled the study of the behaviour of valence quarks. The
first HERA results showed a surprisingly strong rise of $F_2$ as $x$
decreases. An example of such a rise is given in
figure~\ref{fig:f2-over} 
where $F_2$ increases as $x$ decreases, for a
fixed value of $Q^2$ = 15 GeV$^2$.
This increase is the result of the rising gluon density at low
$x$. Note the good agreement between both HERA experiments, H1 and ZEUS,
and also between HERA and the fixed target data.

\subsection{Evolution of $F_2$}

The measurements of $F_2$ as function of $x$ and $Q^2$ can be used to
obtain information about the parton densities in the proton. This is
done by using the pQCD DGLAP evolution equations. One can not
calculate everything from first principles but needs as input from the
experiment the parton densities at a scale $Q^2_0$, usually taken as a
few GeV$^2$, above which pQCD is believed to be applicable. 

There are several groups which perform QCD fits, the most notable are
MRST~\cite{mrst} and CTEQ~\cite{cteq}. They parameterize the $x$
dependence of the parton densities at $Q^2_0$ in the form,
\begin{equation}
xq(x,Q^2) \sim x^{\eta_1}\cdot(1 - x)^{\eta_2}\cdot f_{smooth}(x).
\end{equation}
The free parameters like $\eta_1$ and $\eta_2$ are adjusted to fit the
data for $Q^2>Q^2_0$.
\begin{figure}[htb]
\begin{minipage}{10.5cm}
\epsfig{file=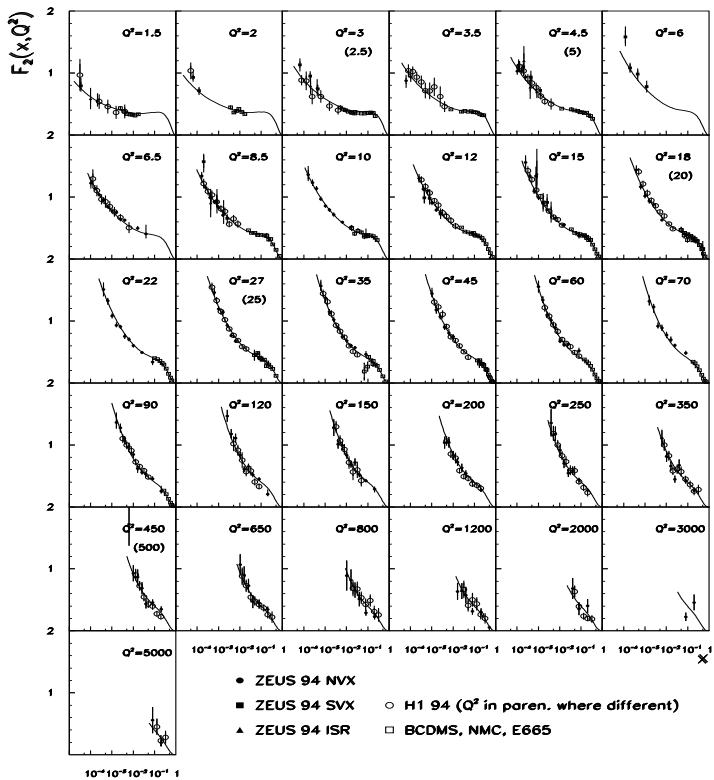,height=10.5cm,width=10.5cm}
\end{minipage}
\hspace*{8mm}
\begin{minipage}{4.5cm}
\vspace*{-1.1cm}
\caption{$F_2$ as function of $x$, for fixed $Q^2$ values (in GeV$^2$) as 
indicated in the figure, for the HERA '94 data together with some
fixed target data. The curves are the result of a NLO QCD fit.}
\label{fig:f2-94}
\end{minipage}
\end{figure}
An example of such an evolution study can be seen in
figure~\ref{fig:f2-94} 
where the $F_2$ data are presented as function
of $x$ for fixed $Q^2$ values. The increase of $F_2$ with decreasing
$x$ is seen over the whole range of measured $Q^2$ values. The pQCD
fits give a good description of the data down to surprisingly low
$Q^2$ values.

\begin{figure}[htb]
\hspace*{-0.4cm}
\begin{minipage}{7.5cm}
\epsfig{file=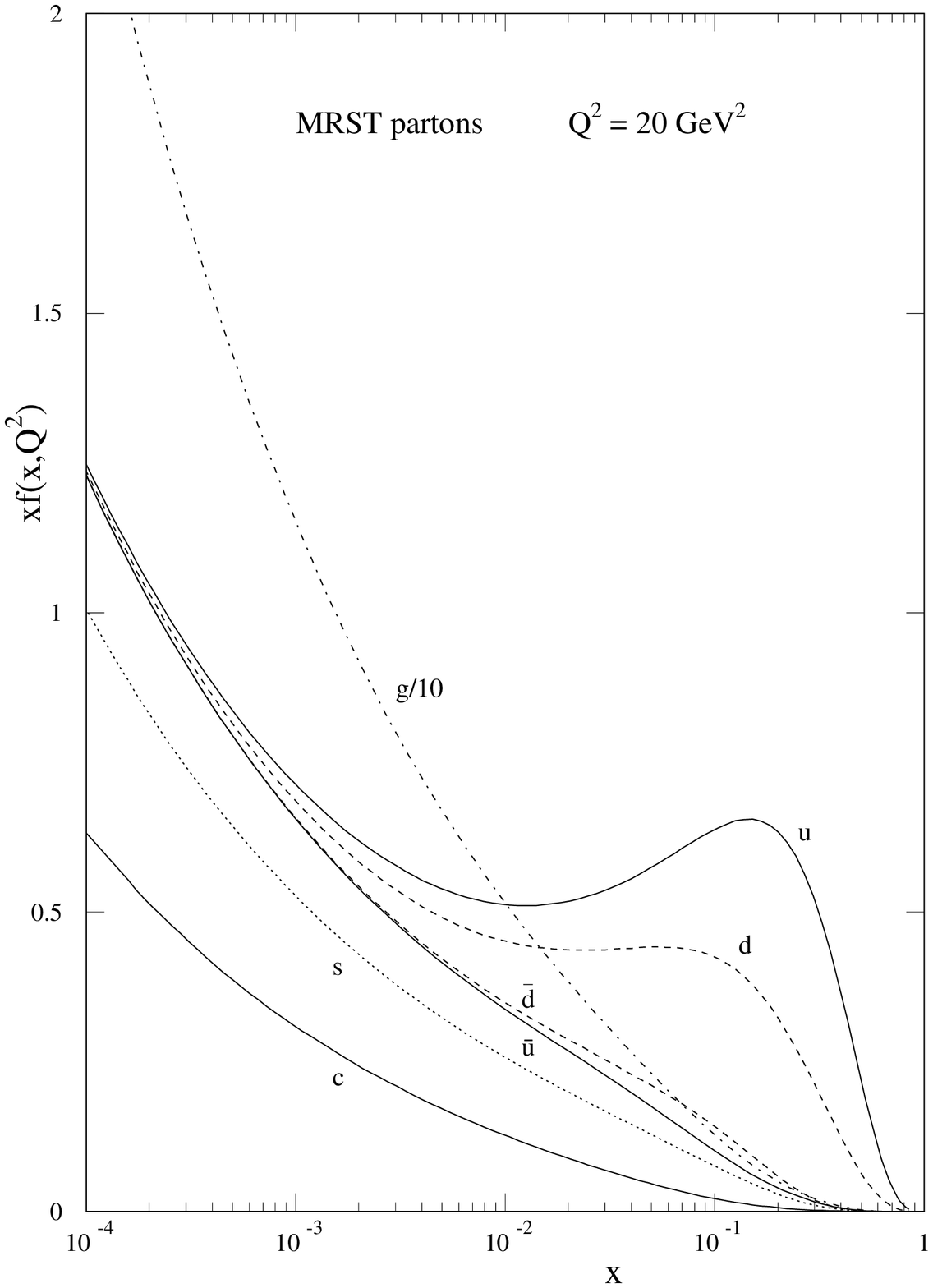,height=11cm}
\vspace*{-1.1cm}
\caption{
Parton density distributions, as function of $x$, of the MRST global
QCD fit, at a scale of $Q^2$ = 20 GeV$^2$.}
\label{fig:mrst-partons}
\end{minipage}
\hspace*{8mm}
\begin{minipage}{7.5cm}
\epsfig{file=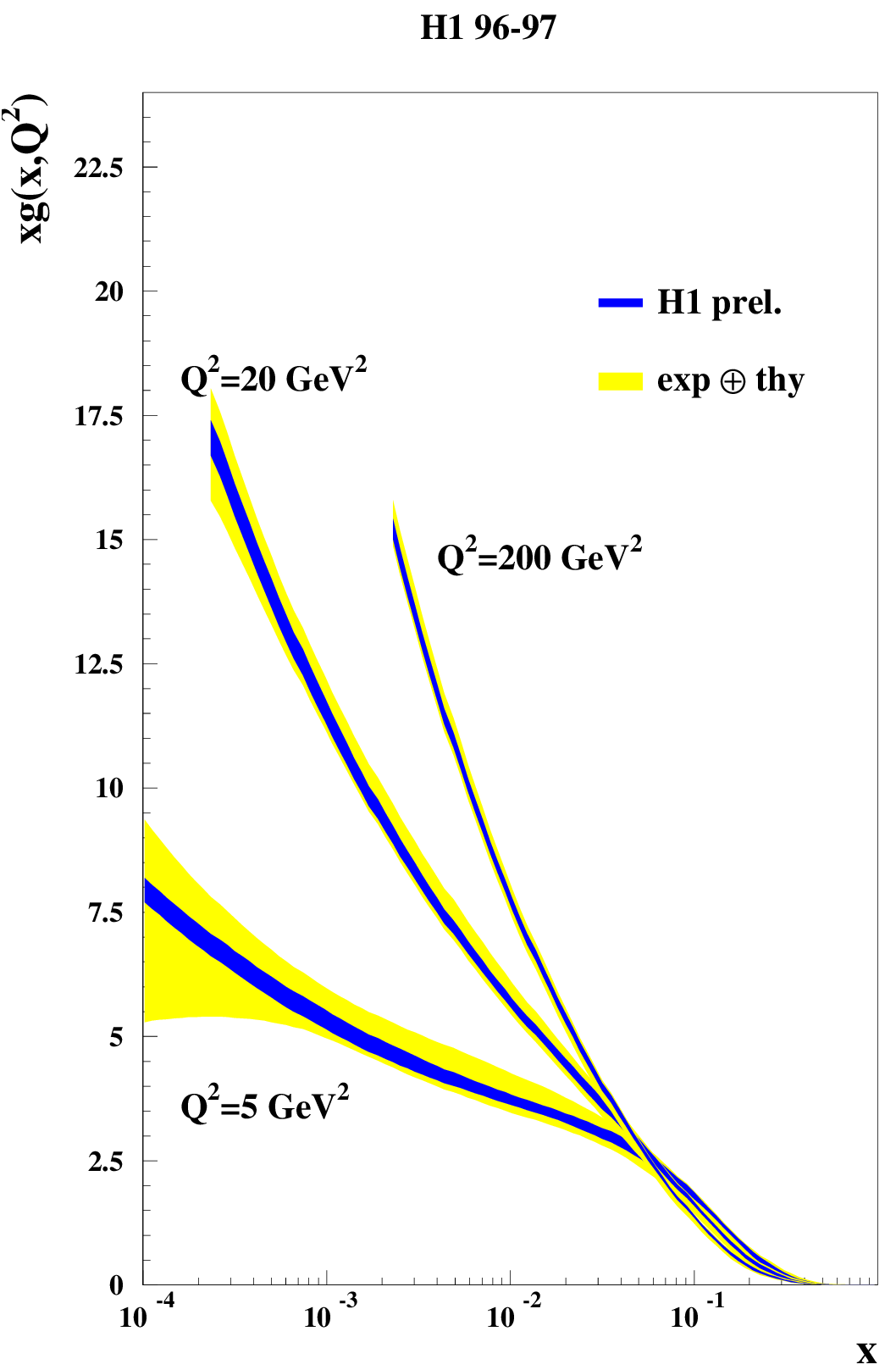,height=11cm}
\vspace*{-1.2cm}
\caption{The gluon density distribution, as function of $x$, at 
$Q^2$ = 5, 20 and 200 GeV$^2$.}
\label{fig:h1-gluon-9697}
\end{minipage}
\end{figure}
The resulting parton densities from the MRST parameterization at $Q^2$
= 20 GeV$^2$ are shown in 
figure~\ref{fig:mrst-partons}. 
One sees the
dominance of the $u$ valence quark at high $x$ and the sharp rise of
the sea quarks at low $x$. In particular, the gluon density at low $x$
rises very sharply and has a value of more than 20 gluons per unit of
rapidity at $x \sim 10^{-4}$. In 
figure~\ref{fig:h1-gluon-9697} 
one sees the extracted gluon density by the H1 experiment~\cite{max}
at three different $Q^2$ values. The density of the gluons at a given
low $x$ increases strongly with $Q^2$.

\subsection{Rise of $F_2$ with decreasing $x$}

\begin{figure}[htb]
\vspace*{-1cm}
\hspace*{-0.5cm}
\begin{minipage}{7.5cm}
\epsfig{file=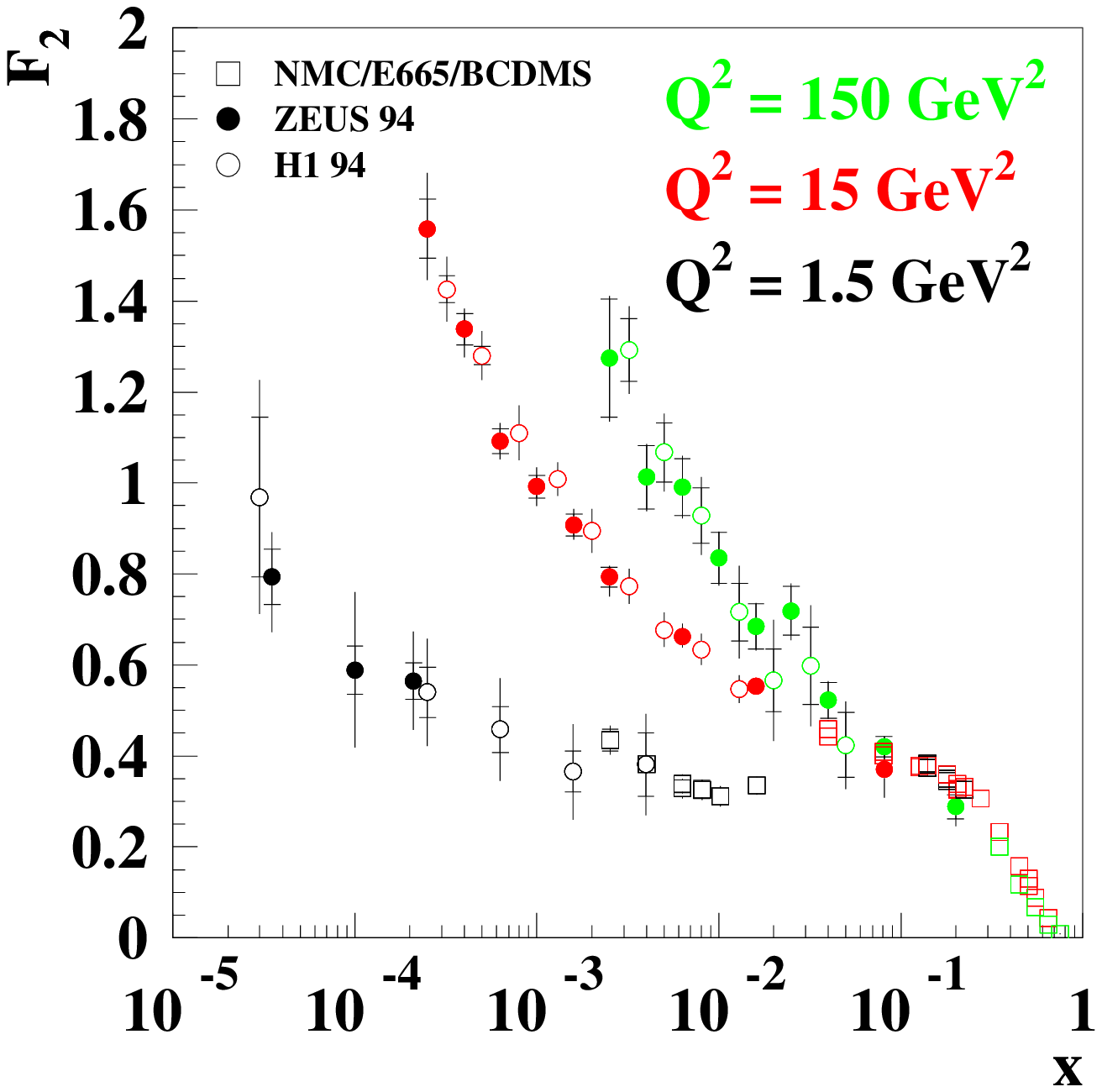,height=7.5cm}
\vspace*{-1.1cm}
\caption{
The proton structure function $F_2$, as function of $x$, for three $Q^2$
values. The higher the $Q^2$, the steeper the distributions.  }
\label{fig:f2-x-rise}
\end{minipage}
\hspace*{5mm}
\begin{minipage}{7.5cm}
\epsfig{file=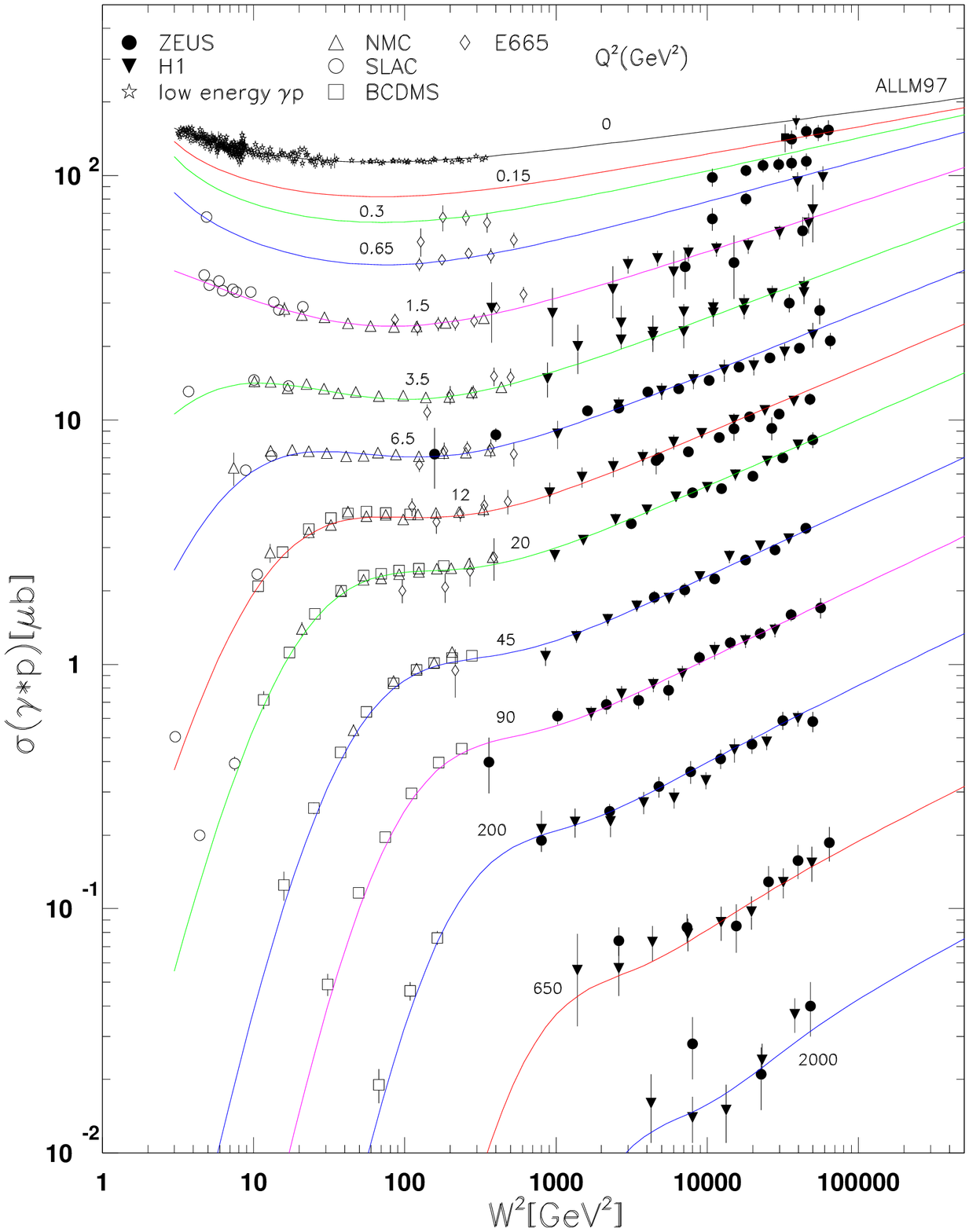,height=11cm}
\vspace*{-1.2cm}
\caption{
The $\gamma^*p$ total cross section $\sigma(\gamma^*p)$ as function of
$W^2$ for fixed values of $Q^2$, including the total photoproduction
cross section ($Q^2$=0). The curves are the results of the ALLM97
parameterization.}
\label{fig:allm97}
\end{minipage}
\end{figure}

The rate of the rise of $F_2$ with decreasing $x$ is $Q^2$
dependent. This can be clearly seen in figure~\ref{fig:f2-x-rise}
where $F_2$ is plotted as a function of $x$ for three $Q^2$
values. The rate of rise decreases as $Q^2$ gets smaller. 
What can we say about the rate of rise? To what can one compare it?
The proton structure function $F_2$ is related to the total $\gamma^* p$
cross section $\sigma_{tot}(\gamma^* p)$,
\begin{equation}
F_2 = \frac{Q^2(1-x)}{4\pi^2\alpha}\frac{Q^2}
{Q^2+4m_p^2x^2}\sigma_{tot}(\gamma^* p) \approx \frac{Q^2}{4\pi^2\alpha}
\sigma_{tot}(\gamma^* p),
\end{equation}
where the approximate sign holds for low $x$.
Since we have a better feeling for the behaviour of the total cross
section with energy, we plot in figure~\ref{fig:allm97} the $F_2$ data
converted to $\sigma_{tot}(\gamma^* p)$ as function of $W^2$ for fixed
values of $Q^2$. For comparison we plot also the total $\gamma p$
cross section. One sees that the shallow $W$ behaviour of the total
$\gamma p$ cross section changes to a steeper behaviour as $Q^2$
increases. The curves are the results of the ALLM97~\cite{allm97}
parameterization (see below) which gives a good description of the
transition seen in the data.

\subsection{The transition region}

The data presented above show a clear change of the $W$ dependence
with $Q^2$. At $Q^2$=0 the processes are dominantly non-perturbative
and the resulting reactions are usually named as `soft' physics. This
domain is well described in the Regge picture. As $Q^2$ increases, the
exchanged photon is expected to shrink and one expects pQCD to take
over. The reactions are said to be `hard'. Where does the transition
from soft to hard physics take place? Is it a smooth or abrupt one? In
the following we describe two parameterizations, one fully based on
the Regge picture while the other combines the Regge approach with a
QCD motivated one.

\subsection{Example of two parameterizations}
\begin{figure}[htb]
\hspace*{-0.7cm}
\epsfig{file=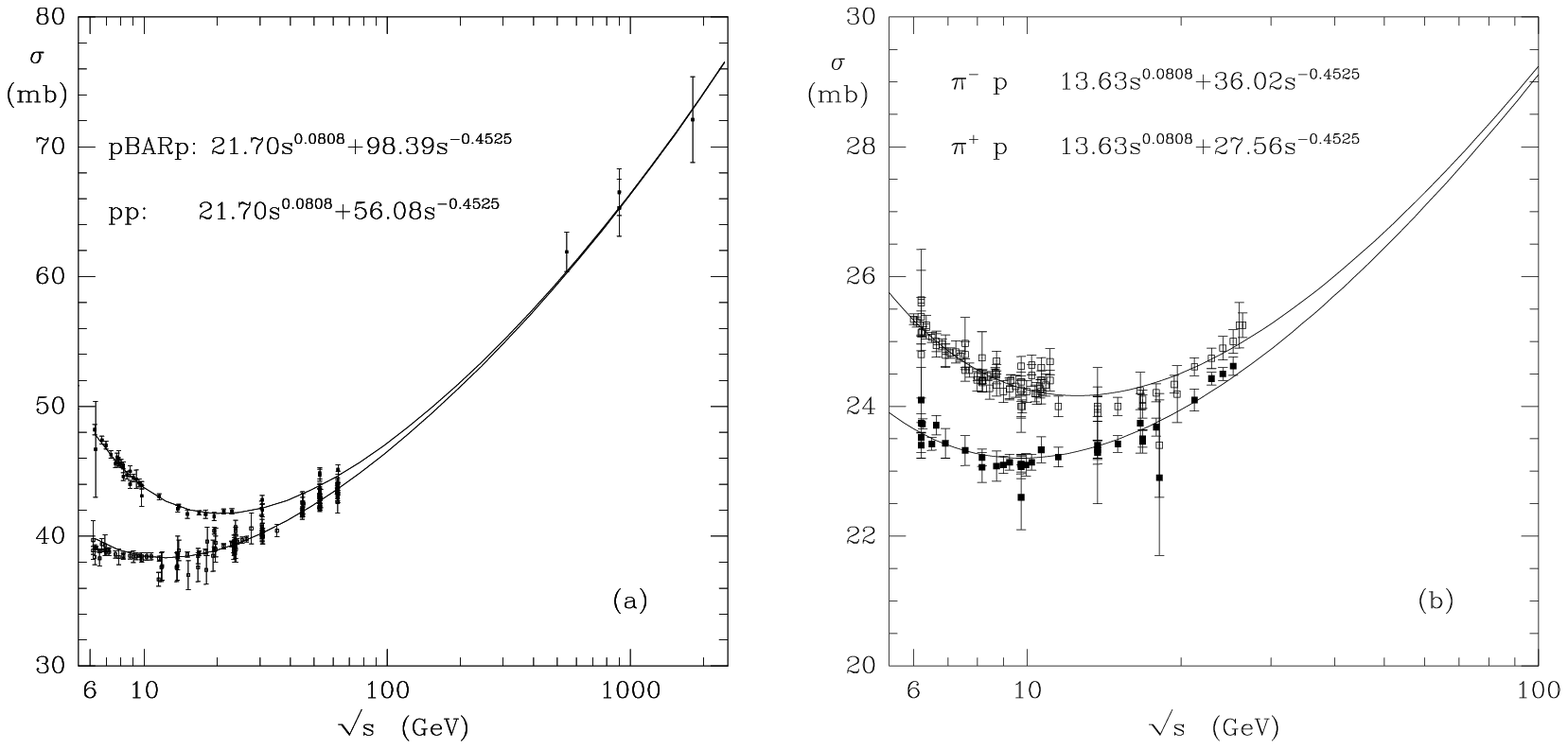,height=5.cm}
\epsfig{file=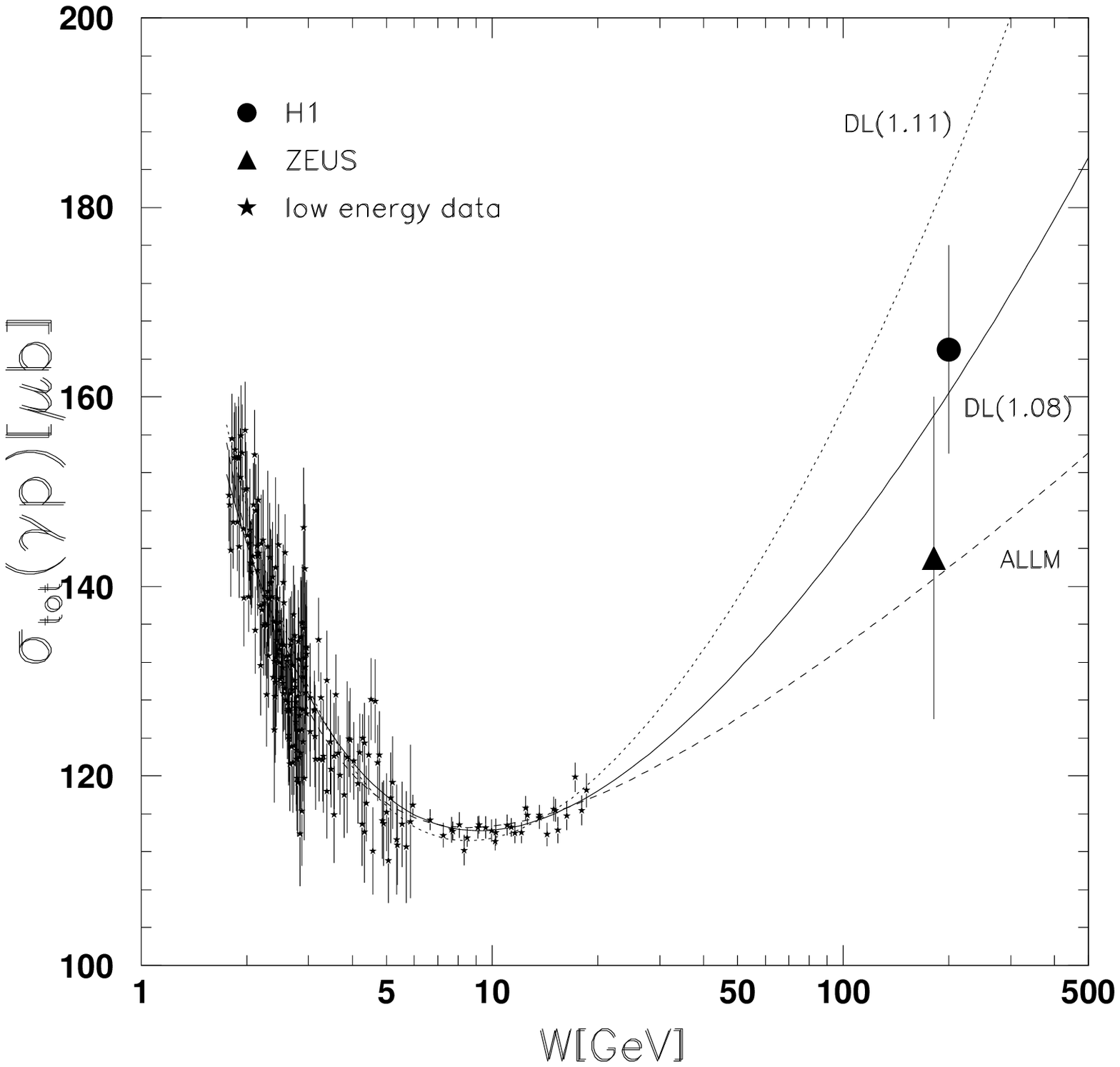,height=5.cm}
\vspace*{-1.1cm}
\caption{
The total cross section data of $\bar{p}p, pp,\pi^\pm p$ and $\gamma
p$ as function of the center of mass energy. The different lines are
the results of parameterizations to these data (see text).}
\label{fig:dl-tot}
\end{figure}

Donnachie and Landshoff (DL)~\cite{dl} succeeded to describe all
existing hadron-proton total cross section data in a simple Regge
picture by using a combination of a Pomeron and a Reggeon exchange,
the former rising slowly while the latter decreasing with energy,
\begin{equation}
\sigma_{tot} = X s^{0.08} + Y s^{-0.45},
\end{equation}
where $s$ is the square of the total center of mass energy. The two
numerical parameter, related to the intercepts $\alpha(0)$ of the
Pomeron and Reggeon trajectories, respectively, are the result of
fitting this simple expression to all available data, some of which
are shown in the first two plots in figure~\ref{fig:dl-tot}. These
parameters give also a good description of the total $\gamma p$ cross
section data which were not used in the fit and are also shown on the
right hand side of the figure.
Donnachie and Landshoff wanted to extend this picture also to virtual
photons~\cite{dlq2} (for $Q^2<$10 GeV$^2$), keeping the power of
$W^2$, which is related to the Pomeron intercept, fixed with
$Q^2$. Their motivation was to see what is the expected contribution
from non-perturbative physics, or soft physics as we called it above,
at higher $Q^2$.

The other example is that of Abramowicz, Levin, Levy, Maor
(ALLM)~\cite{allm}, which was updated by Abramowicz and Levy
(ALLM97)~\cite{allm97}. This parameterization uses a Regge motivated
approach at low $x$ together with a QCD motivated one at high $x$ to
parameterize the whole ($x,Q^2$) phase space, fitting all existing
$F_2$ data. This parameterization uses a so-called interplay of soft
and hard physics (see~\cite{interplay}) . 

\begin{figure}[htb]
\begin{minipage}{7.5cm}
\hspace*{-1cm}
\epsfig{file=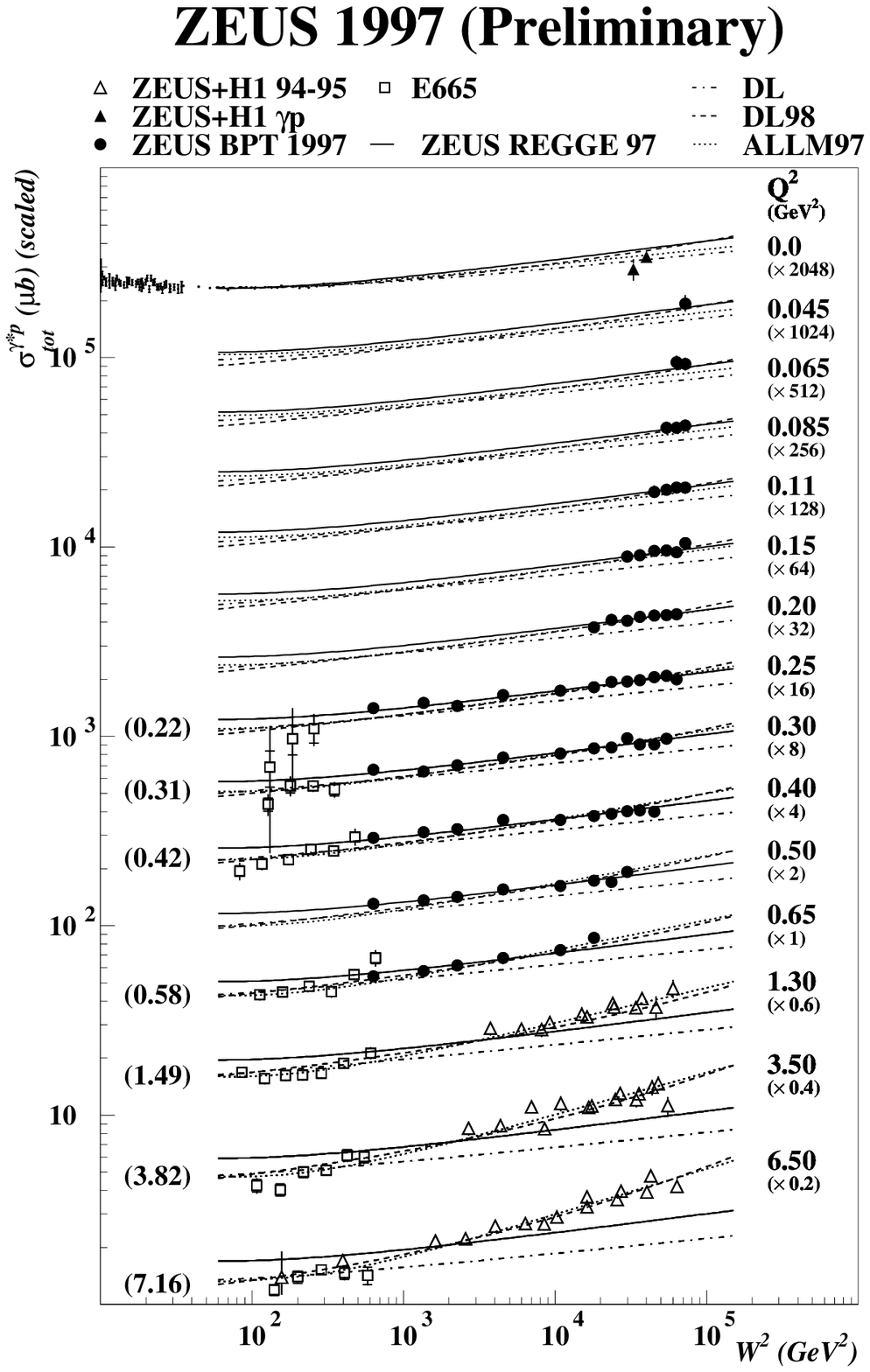,height=12.5cm}
\vspace*{-1.cm}
\caption{
The $\gamma^* p$ total cross section, as function of $W^2$, in bins of
$Q^2$ for HERA data and that of E665, compared with some
parameterizations.}
\label{fig:sig97}
\end{minipage}
\hspace*{8mm}
\begin{minipage}{7.5cm}
\epsfig{file=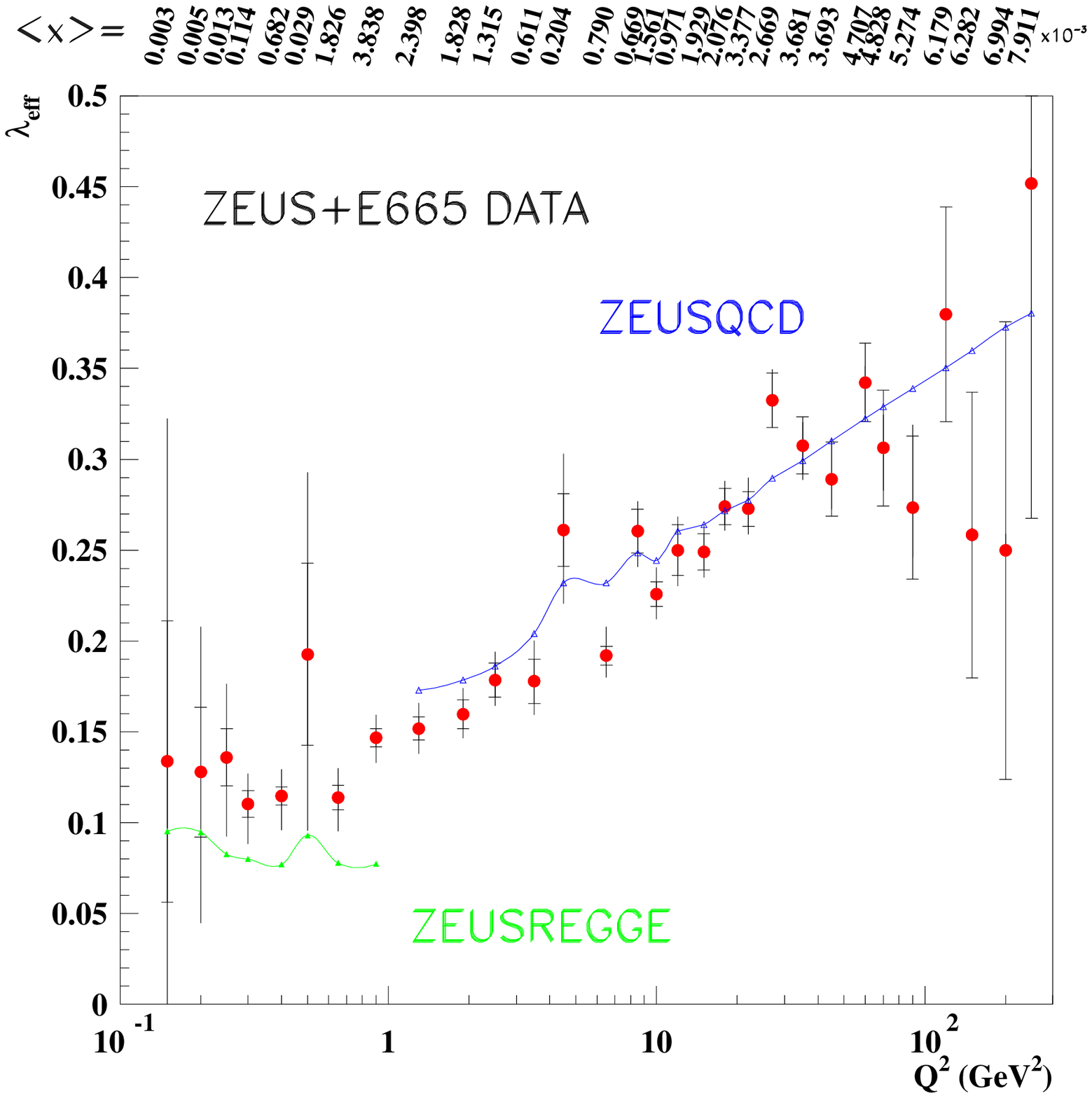,height=5.5cm}
\epsfig{file=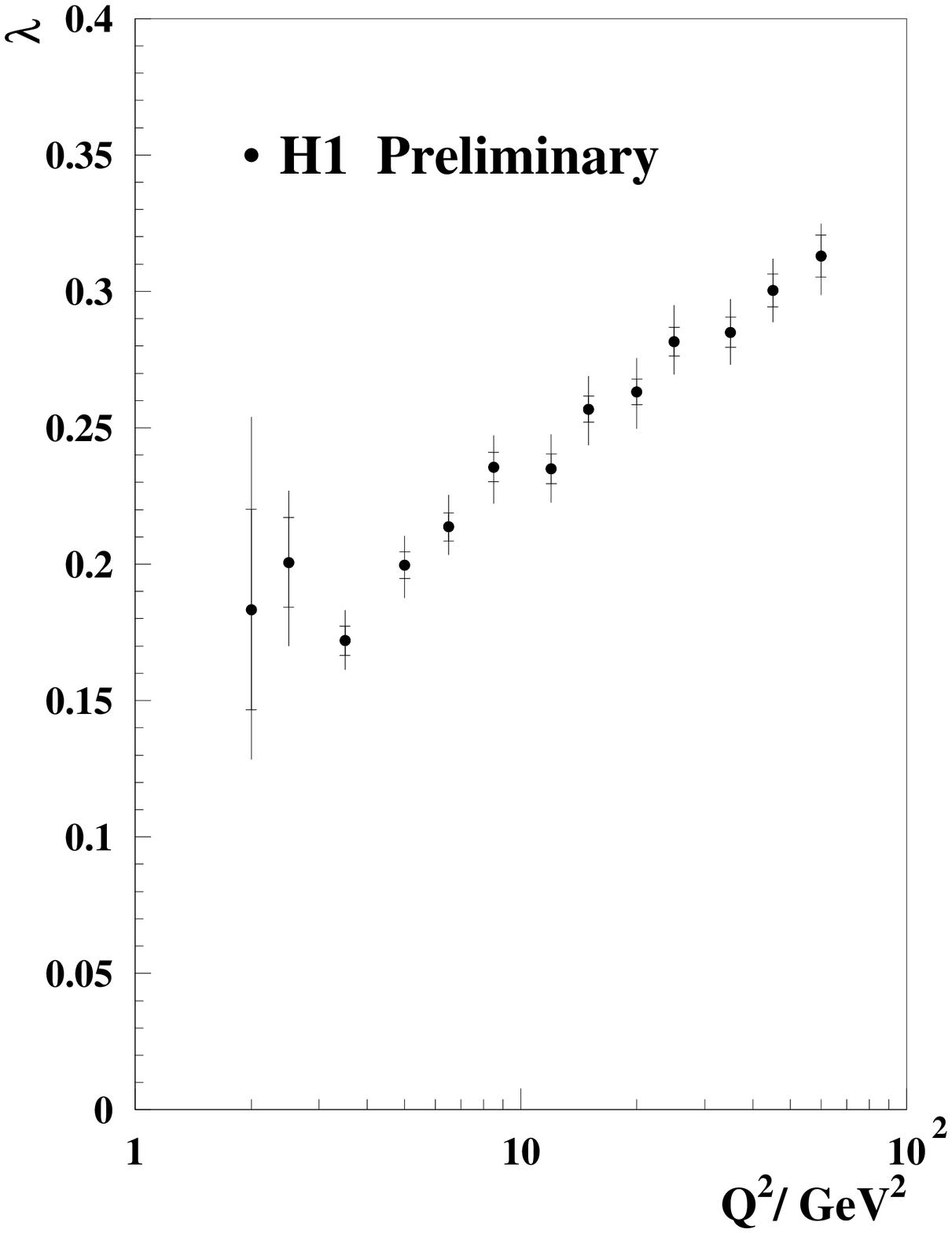,height=7cm}
\vspace*{-1.cm}
\caption{The $Q^2$ dependence of the parameter $\lambda$, determined from 
a fit of the form $F_2 \sim x^{-\lambda}$ at fixed $Q^2$ values.}
\label{fig:lam}
\end{minipage}
\end{figure}
The two parameterizations are compared~\cite{amelung} to the low $Q^2$
HERA data together with that of the fixed target E665 experiment in
figure~\ref{fig:sig97}. Here one sees again how the cross section
changes from a $(W^2)^{0.08}$ behaviour at very low $Q^2$ to a
$(W^2)^{0.2-0.4}$ as $Q^2$ increases. The simple DL parameterization
as implemented by ZEUS (ZEUSREGGE in the figure) fails to describe the
data above $Q^2 \sim$ 1 GeV$^2$. ALLM97 describes the data well in the
whole region. DL98~\cite{dl98}, which adds to the soft Pomeron an
additional hard Pomeron, can also describe the data, but loses the
simplicity of the original DL one.

One can quantify the change in the rate of increase by using the
parameter $\lambda$. Since $\sigma_{tot} \sim (W^2)^{\alpha(0)-1}$
this implies that $F_2 \sim x^{-\lambda}$. The fitted value of
$\lambda$ as function of $Q^2$ is shown in figure~\ref{fig:lam} for
the ZEUS~\cite{zlam} (upper) and the H1~\cite{h1lam} (lower)
experiments. One sees a clear increase of $\lambda$ with $Q^2$ which
cannot be reproduced by the simple Regge picture but needs an approach
in which there is the interplay of soft and hard physics~\cite{interplay}.

\subsection{What have we learned about the structure of the proton?}

Let us summarize what we have learned so far about the structure of
the proton. 
\begin{itemize}
\item The density of partons in the proton increases with
decreasing $x$. 
\item The rate of increase is $Q^2$ dependent; at high $Q^2$
the increase follows the expectations from the pQCD hard physics while
at low $Q^2$ the rate is described by the soft physics behaviour
expected by the Regge phenomenology. 
\item Though there seems to be a
transition in the region of $Q^2 \sim$ 1-2 GeV$^2$, there is an
interplay between the soft and hard physics in both regions.
\end{itemize}

\section{THE STRUCTURE OF THE PHOTON}

In this part we will describe what is presently known about the
structure of the photon, both from $e^+e^-$ experiments as well as
from HERA.

\subsection{Photon structure from $e^+e^-$}

The hadronic structure function of the photon, $F_2^\gamma$, was
measured in $e^+e^-$ collisions which can be interpreted as depicted
in figure~\ref{fig:dis-photon}. A highly virtual $\gamma^*$ with
large $Q^2$ probes a quasi-real $\gamma$ with $P^2 \approx$ 0.

\begin{figure}[htb]
\begin{minipage}{7.5cm}
\hspace*{-5mm}
\epsfig{file=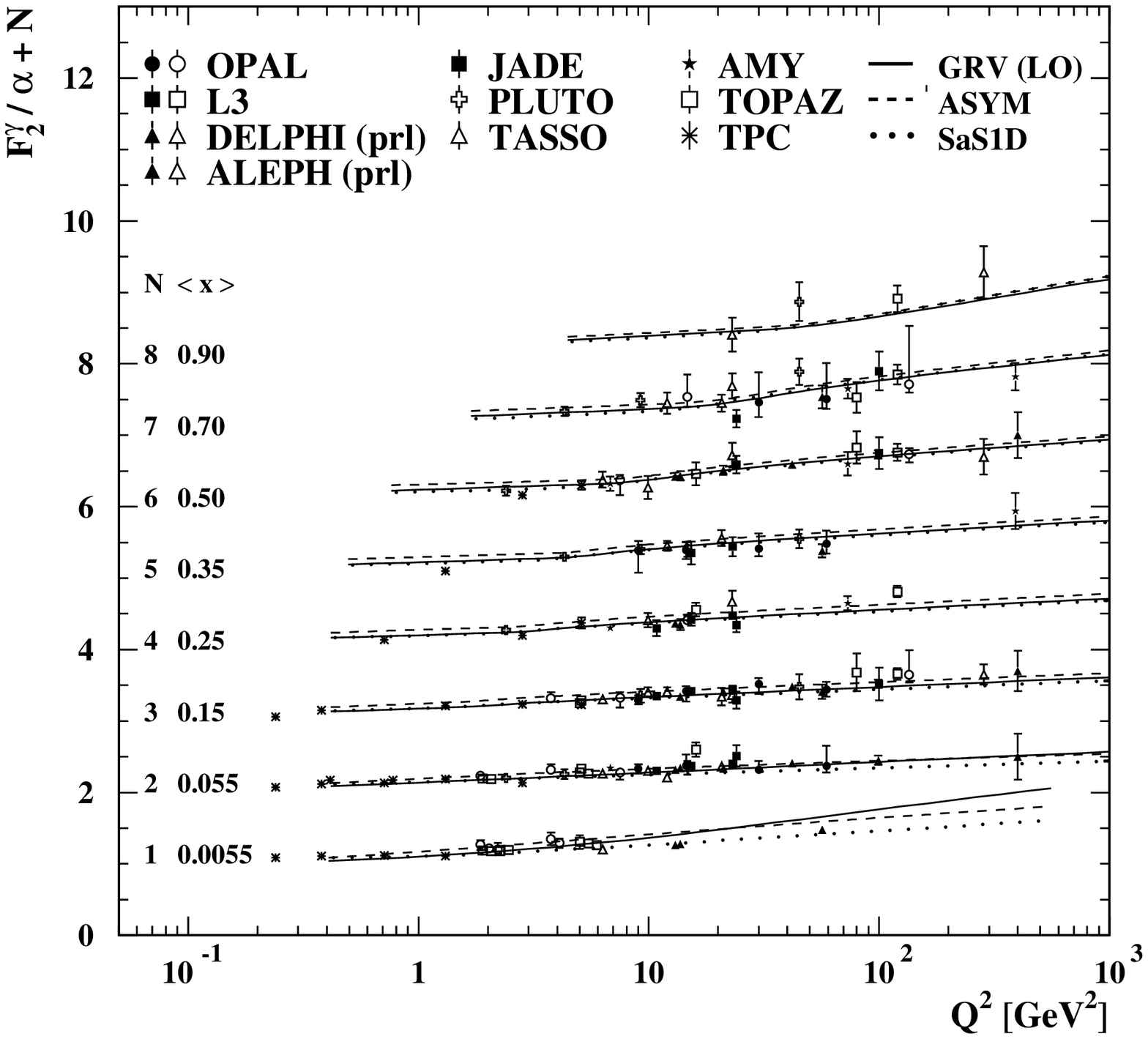,height=7.5cm}
\vspace*{-1.cm}
\caption{
The photon structure function $F_2^\gamma$, as function of $Q^2$, for
average $x$ values as given in the figure. The curves are the
expectations of different parameterizations of parton distributions in
the photon.}
\label{fig:gscaling}
\end{minipage}
\hspace*{4mm}
\begin{minipage}{7.5cm}
\vspace*{-7mm}
\hspace*{-8mm}
\epsfig{file=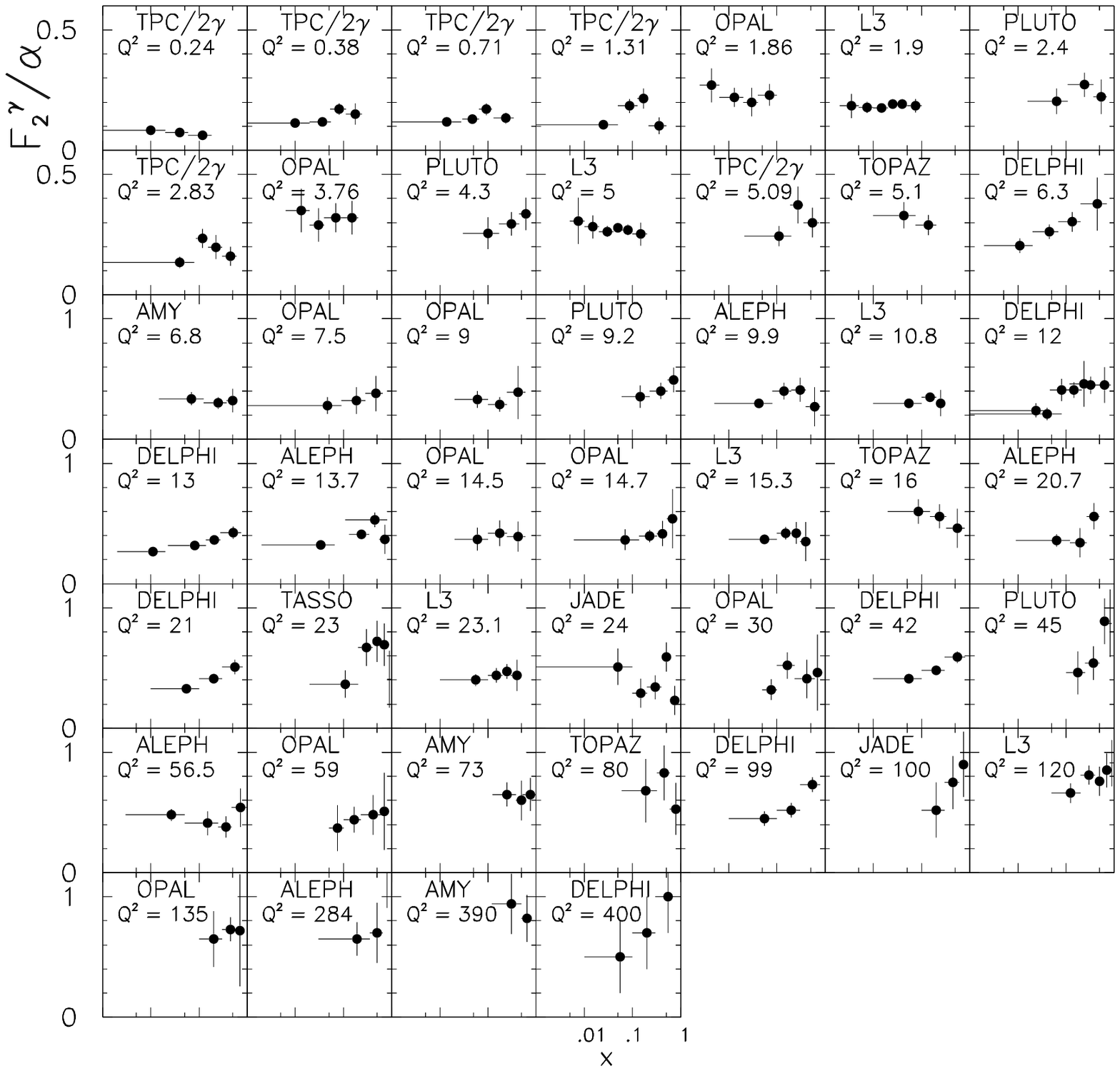,height=9.5cm}
\vspace*{-1.5cm}
\hspace*{12mm}
\caption{
$F_2^\gamma$, as function of $x$, for fixed $Q^2$ as given in the
figure. The data points have been taken from the numerical tables 
in~\cite{nisius}.}  
\label{fig:f2g}
\end{minipage}
\end{figure}

The measurements of $F_2^\gamma$ showed a different behaviour than
that of the proton structure function. From the $Q^2$ dependence,
shown in figure~\ref{fig:gscaling}~\cite{nisius}, one sees positive
scaling violation for all $x$.
This different behaviour can be understood as coming from an
additional splitting to the ones present in the proton case (see
figure~\ref{fig:splitting}). In the photon case, the photon can split
into a $q\bar{q}$ pair, $\gamma \to q\bar{q}$. The contribution
resulting from this splitting, called the `box diagram', causes
positive scaling violation for all $x$. In addition, and again
contrary to the proton case, it also causes the photon structure
function to be large for high $x$ values, as can be seen in
figure~\ref{fig:f2g} where $F_2^\gamma$ is plotted as function of $x$
for fixed $Q^2$ values. From this figure one can also see that there
exist very little data in the low $x$ region.

\subsection{Photon structure from HERA}

\begin{figure}[htb]
\begin{minipage}{7.5cm}
\epsfig{file=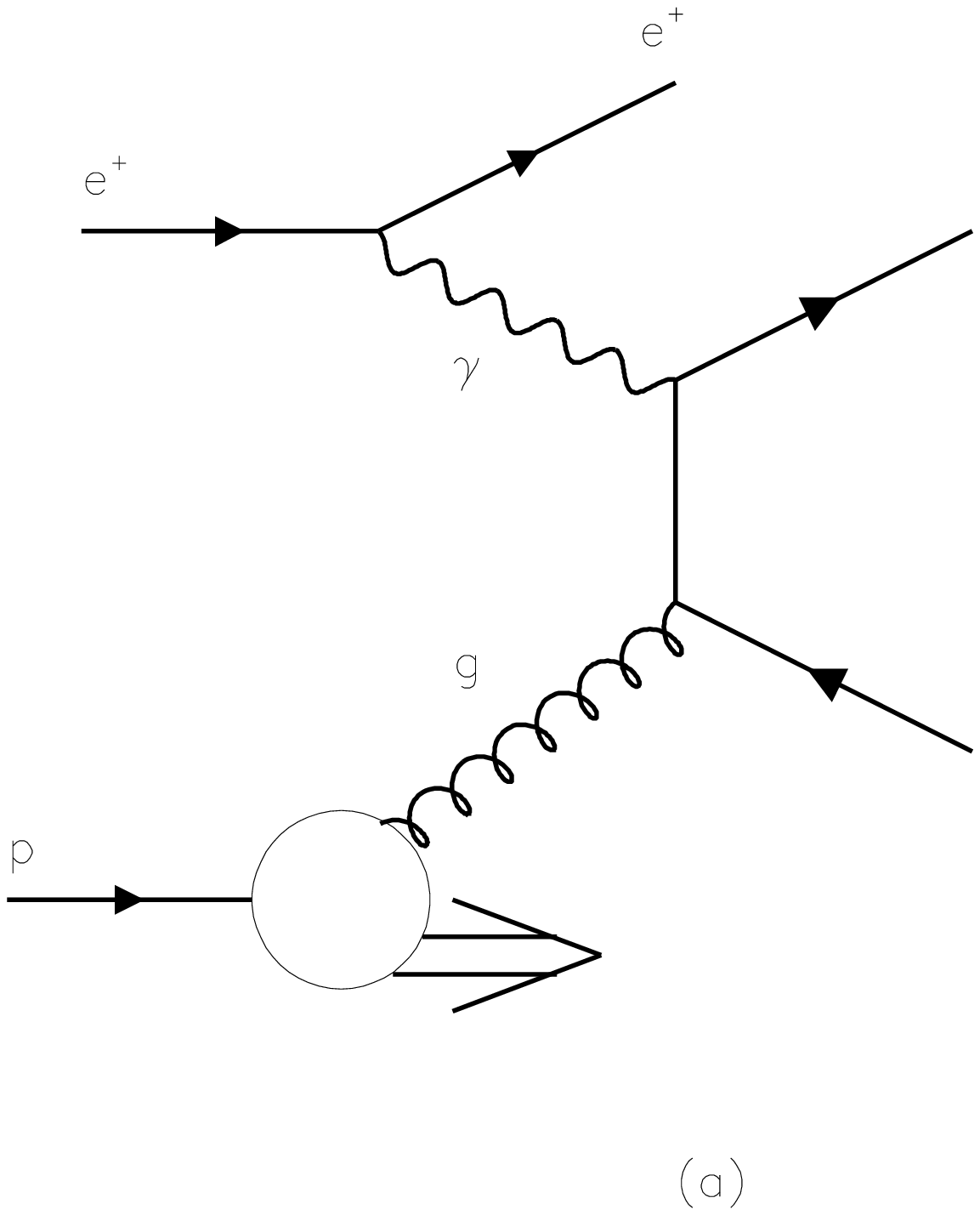,height=3cm}
\epsfig{file=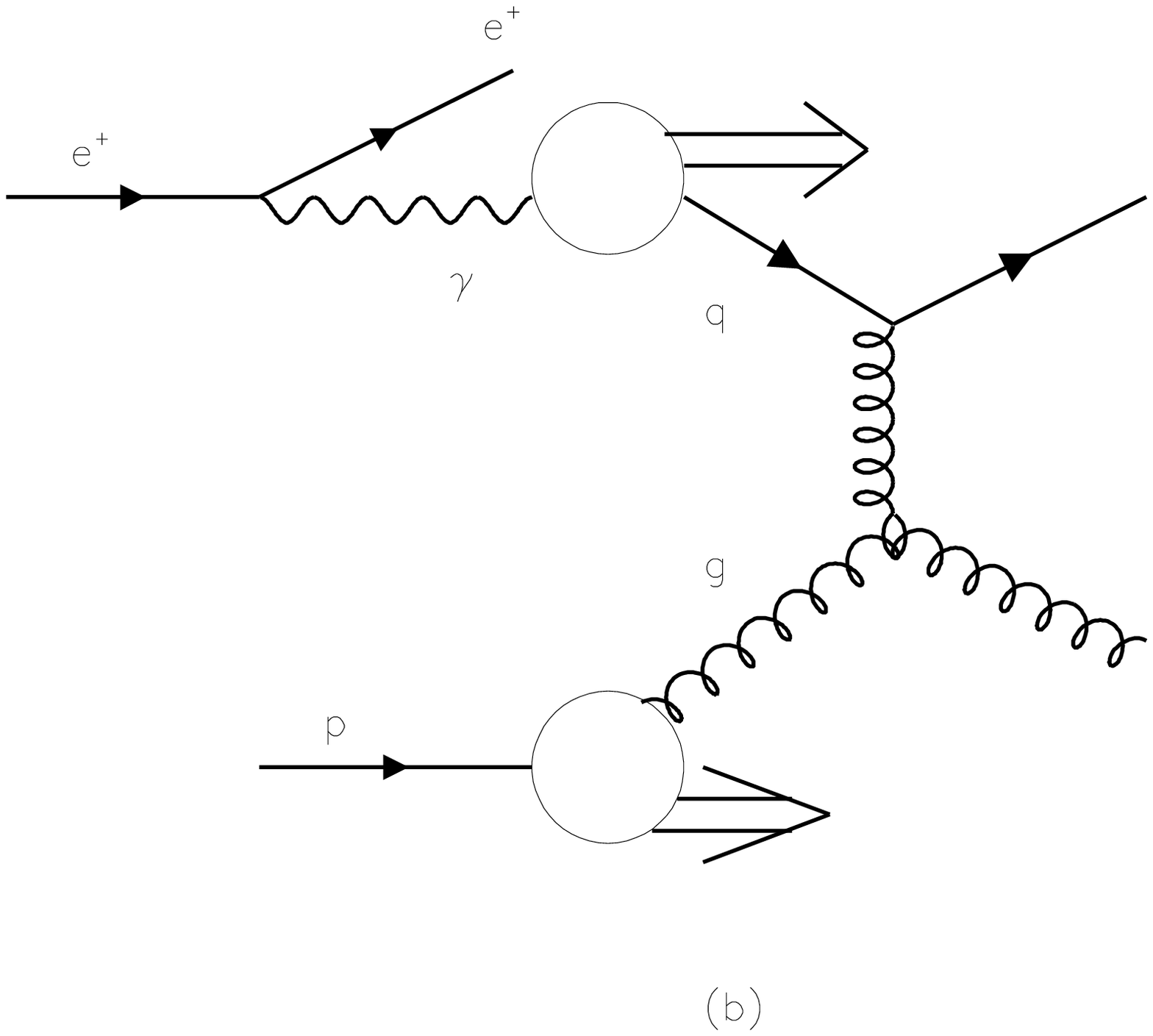,height=3cm}
\caption{Examples of leading order QCD (a) `direct' and (b) `resolved' 
dijet production diagrams.}
\label{fig:direct-resolved}
\end{minipage} 
\hspace*{5mm}
\begin{minipage}{7.5cm}
\vspace*{-1.5cm}
\epsfig{file=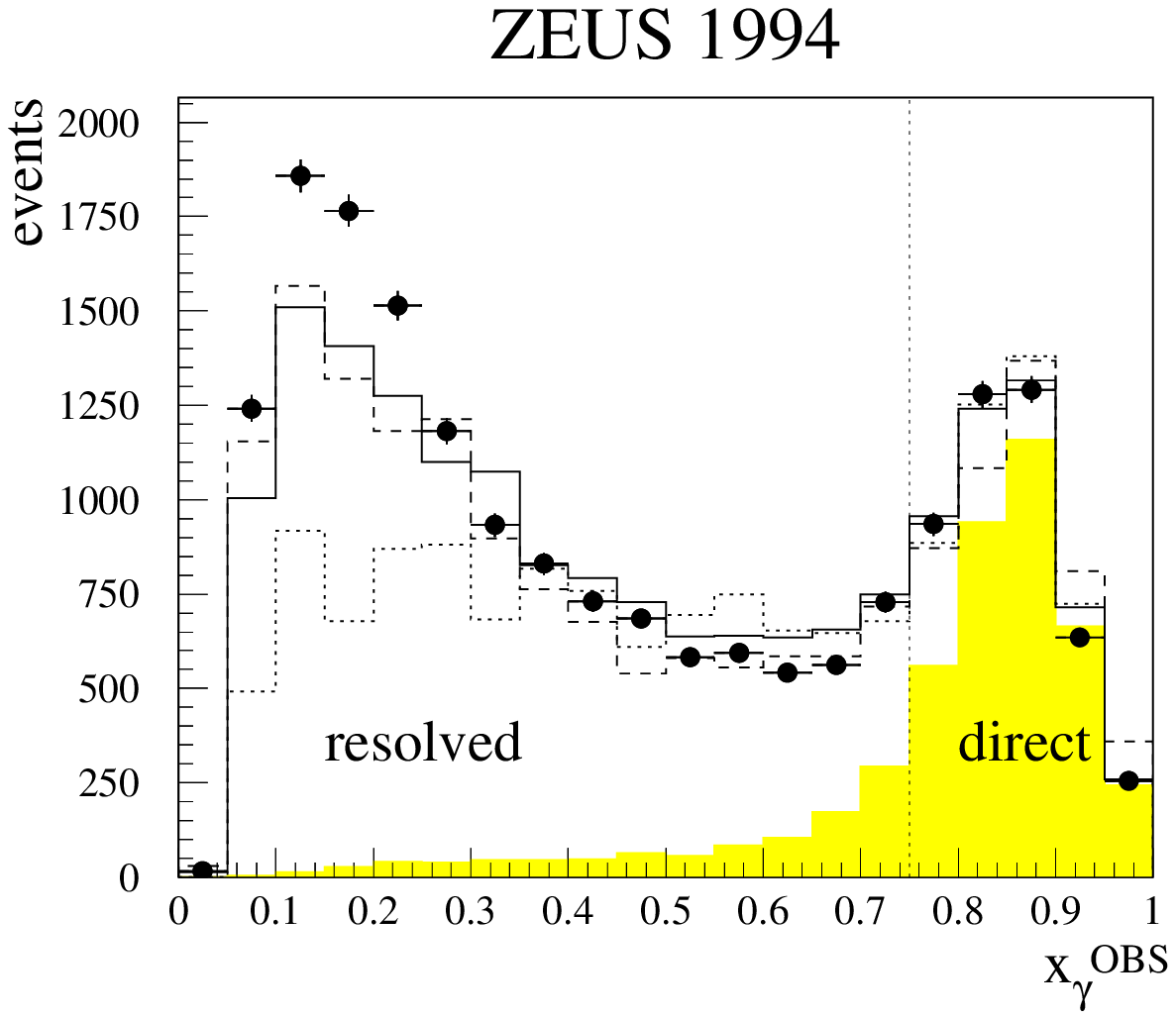,height=5.5cm}
\vspace*{-1cm}
\caption{The $x_\gamma^{obs}$ distribution, as obtained from photoproduction 
of dijet events. The shaded area are the expectations of the distribution 
of this variable from the generation of direct photon events. The dotted 
vertical line is the border of an operational definition of direct and 
resolved photon events.}
\label{fig:xg}
\end{minipage} 
\end{figure}

At HERA, the structure of the photon can be studied by selecting
events in which the exchanged photon is quasi-real and the probe is
provided by a large transverse momentum parton from the proton. The
probed photon can participate in the process in two ways. In one, the
interaction takes place before it fluctuates into a $q\bar{q}$ pair
and thus the whole of the photon participates in the interaction. Such
a process is called a `direct' photon interaction. In the other case,
the photon first fluctuates into partons and only one of these partons
participates in the interaction while the rest continue as the photon
remnant. This process is said to be a `resolved' photon
interaction. An example of leading order diagrams describing dijet
photoproduction for the two processes is shown in
figure~\ref{fig:direct-resolved}.
If one defines a variable $x_\gamma$ as the fraction of the photon
momentum taking part in a dijet process, we expect $x_\gamma
\sim$ 1 in the direct case, while $x_\gamma \ll$ 1 in the resolved photon 
interaction. These two processes are clearly seen in
figure~\ref{fig:xg} where the $x_\gamma$ distribution shows
a two peak structure, one coming from the direct photon and the other
from the resolved photon interactions~\cite{xg}.

\begin{figure}[hb]
\begin{minipage}{8.5cm}
\vspace*{-7mm}
\hspace*{-5mm}
\epsfig{file=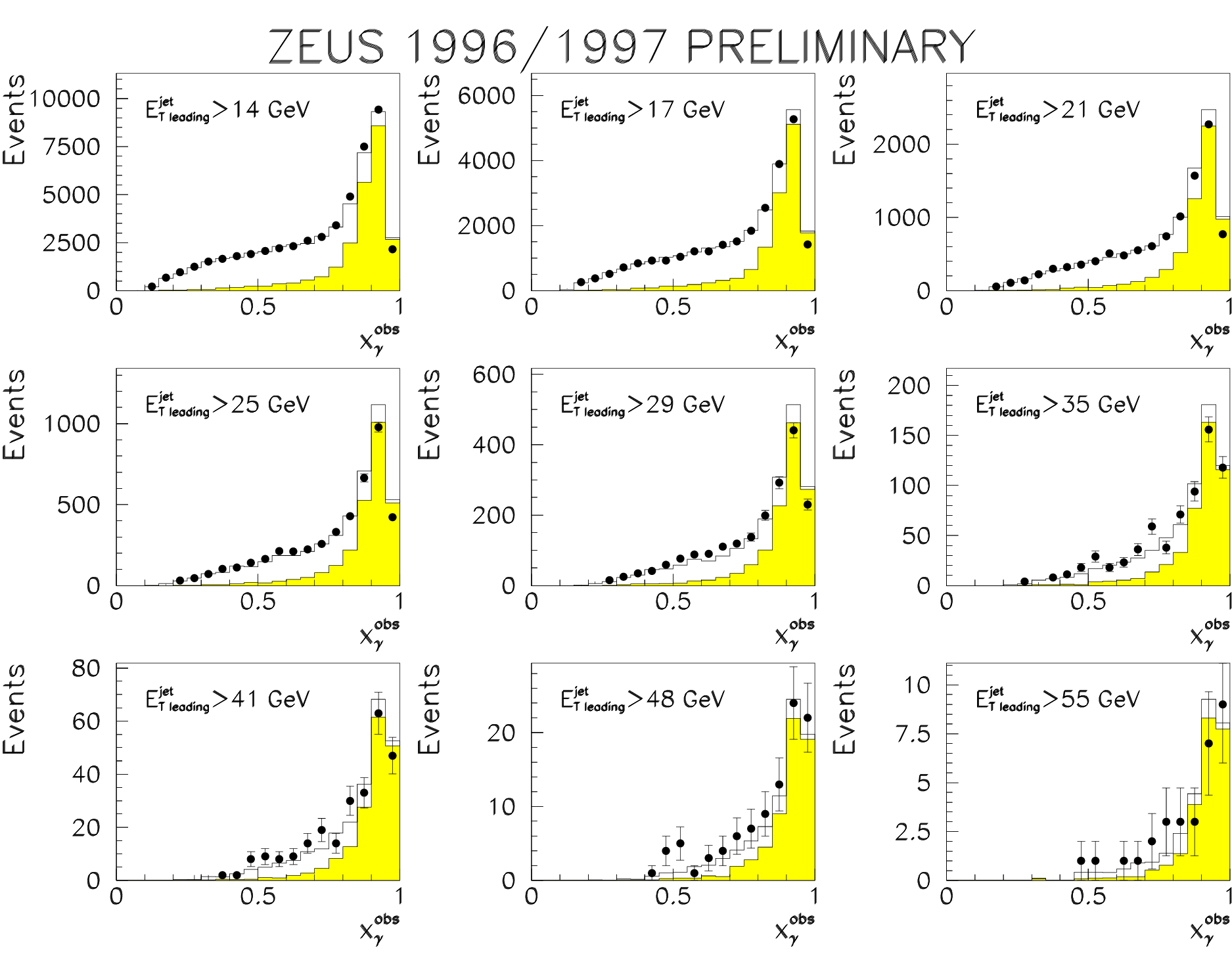,height=6.3cm}
\vspace*{-1.cm}
\caption{
Distributions of $x_\gamma^{obs}$, for different thresholds on the
highest transverse energy jet. The open histogram is the prediction of
the MC, and the shaded part is the direct photon component of the MC.}
\label{fig:f2g-hera}
\end{minipage}
\hspace*{8mm}
\begin{minipage}{6.5cm}
\epsfig{file=gh1.ps,height=5.cm}
\vspace*{-1.cm}
\caption{
Comparison of the photon gluon density, determined from di-jet photoproduction
events taken in 1996, with earlier measurements of H1. The curves are the 
expectations of different parameterizations.}
\label{fig:gh1}
\end{minipage}
\end{figure}

One way of obtaining information about $F_2^\gamma$ from HERA is to
measure the dijet photoproduction as function of $x_\gamma$ and to
subtract the contribution coming from the direct photon
reactions. This is shown in figure~\ref{fig:f2g-hera}, 
where the measurements are presented at fixed values of the hard
scale, which is taken as the highest transverse energy
jet~\cite{f2ghera}.  One can go one step further by assuming leading
order QCD and Monte Carlo (MC) models to extract the effective parton
densities in the photon. An example of the extracted gluon density in
the photon~\cite{gh1} is shown in figure~\ref{fig:gh1}. The gluon
density increases with decreasing $x$, a similar behaviour to that of
the gluon density in the proton. The data have the potential of
differentiating between different parameterization of the parton
densities in the photon, as can be seen in the same figure.

\subsection{Virtual photons at HERA}

\begin{figure}[htb]
\begin{minipage}{7.5cm}
\hspace*{-5mm}
\epsfig{file=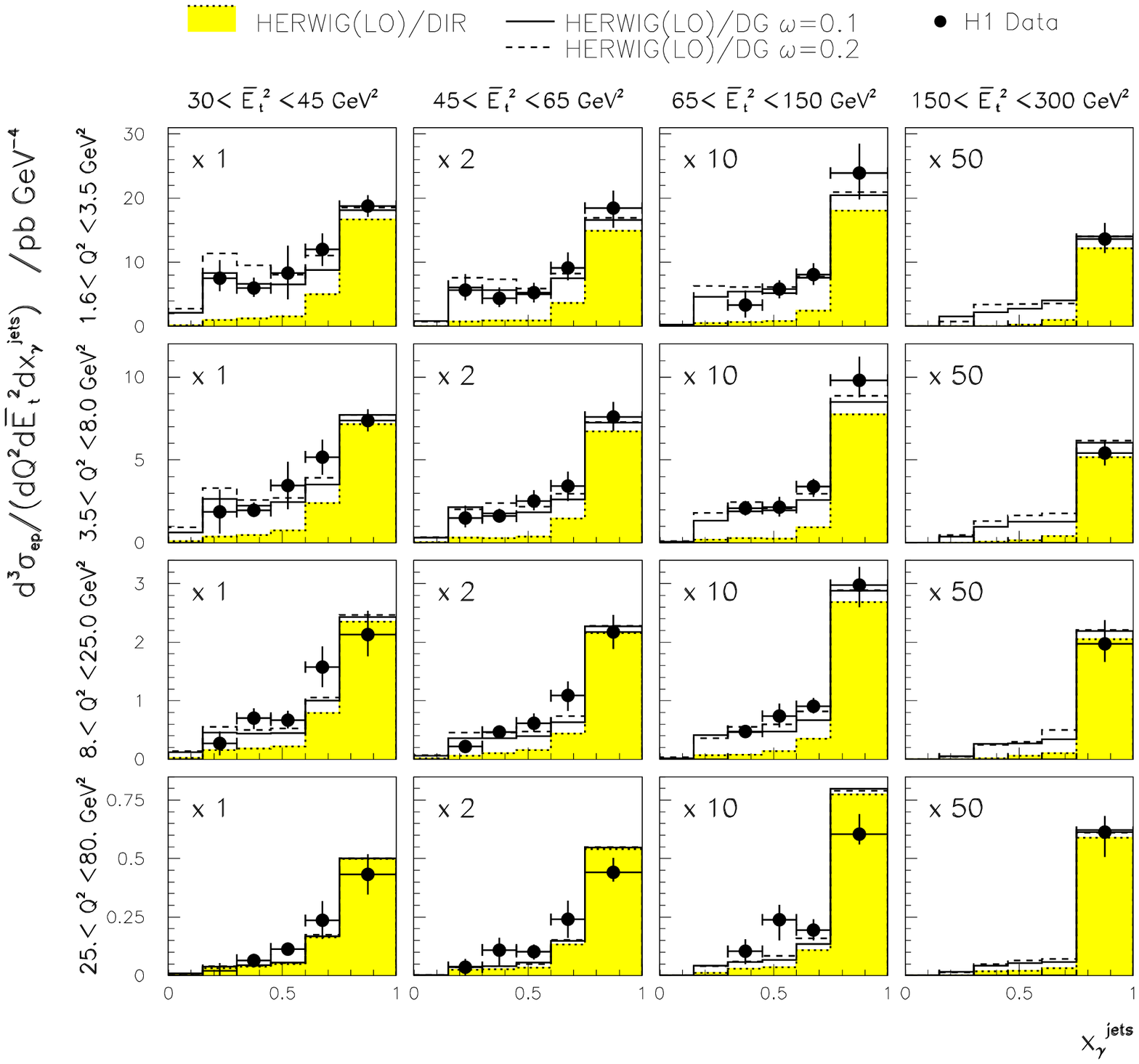,height=8cm}
\vspace*{-1.5cm}
\caption{
The differential dijet cross section, as function of $x_\gamma^{jets}$,
for different scales as indicated in the figure. The open histogram is
the prediction of the MC and the shaded part is the direct photon
component of the MC.}
\label{fig:virt-h1}
\end{minipage}
\hspace*{8mm}
\begin{minipage}{7.5cm}
\epsfig{file=virt-zeus.ps,height=8cm}
\vspace*{-1.cm}
\caption
{The ratio of the resolved to direct photon component, as function of the 
virtuality $Q^2$ (in GeV$^2$) of the photon.} 
\label{fig:virt-zeus}
\end{minipage}
\end{figure}

One can study the structure of virtual photons in a similar way as
described above. In this case, the $Q^2$ of the virtual photon has to
be much smaller than the transverse energy squared of the jet,
$E_t^2$, which provides the hard scale of the probe. Such a
study~\cite{virth1} is presented in figure~\ref{fig:virt-h1}, where
the dijet cross section is plotted as function of $x_\gamma$ for
different regions in $Q^2$ and $E_t^2$. One sees a clear excess over
the expectation of direct photon reactions, indicating that virtual
photons also have a resolved part.
This fact can also been seen in figure~\ref{fig:virt-zeus} where the
ratio of resolved to direct photon interactions is plotted as function
of the virtuality $Q^2$ of the probed photon~\cite{virtz}. One sees
that although the ratio decreases with $Q^2$, it remains
non-zero even at relatively high $Q^2$ values.

\subsection{Virtual photons at LEP}

The study of the structure of virtual photons in $e^+e^-$ reactions
was dormant for more than 15 years following the measurement done by
the PLUTO collaboration~\cite{pluto}. Recently, however, the L3
collaboration at LEP~\cite{virtl3} measured the structure function of
photons with a virtuality of 3.7 GeV$^2$, using as probes photons with
a virtuality of 120 GeV$^2$.
\begin{figure}[htb]
\begin{minipage}{7.5cm}
\hspace*{-1cm}
\epsfig{file=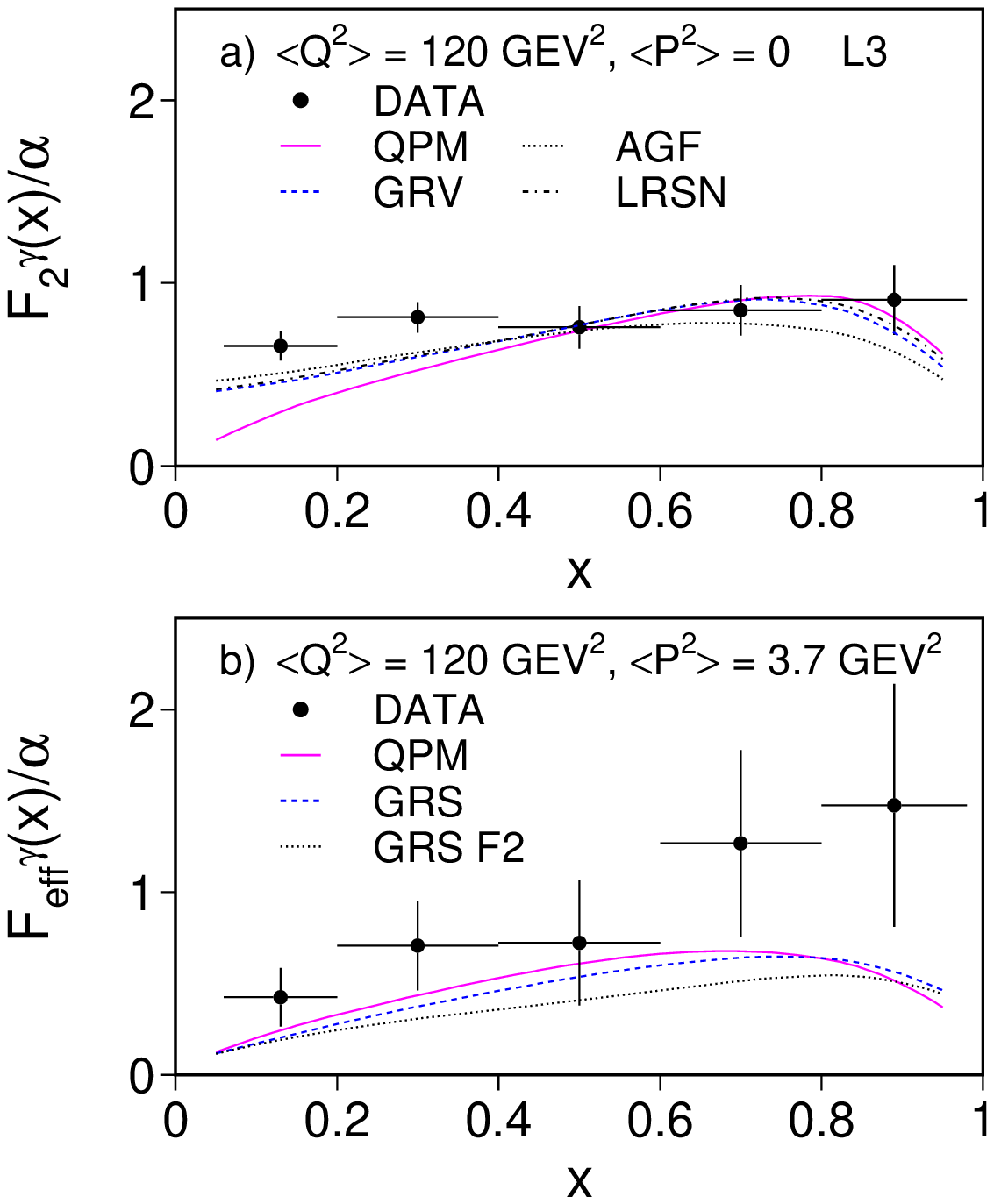,height=10cm}
\vspace*{-1.cm}
\caption{
The effective photon structure function from L3 for (a) a quasi-real photon 
target and (b) a 3.7 GeV$^2$ photon target. Both photons were probed at a 
scale of 120 GeV$^2$.}
\label{fig:virtlep1}
\end{minipage}
\hspace*{3mm}
\begin{minipage}{7.5cm}
\epsfig{file=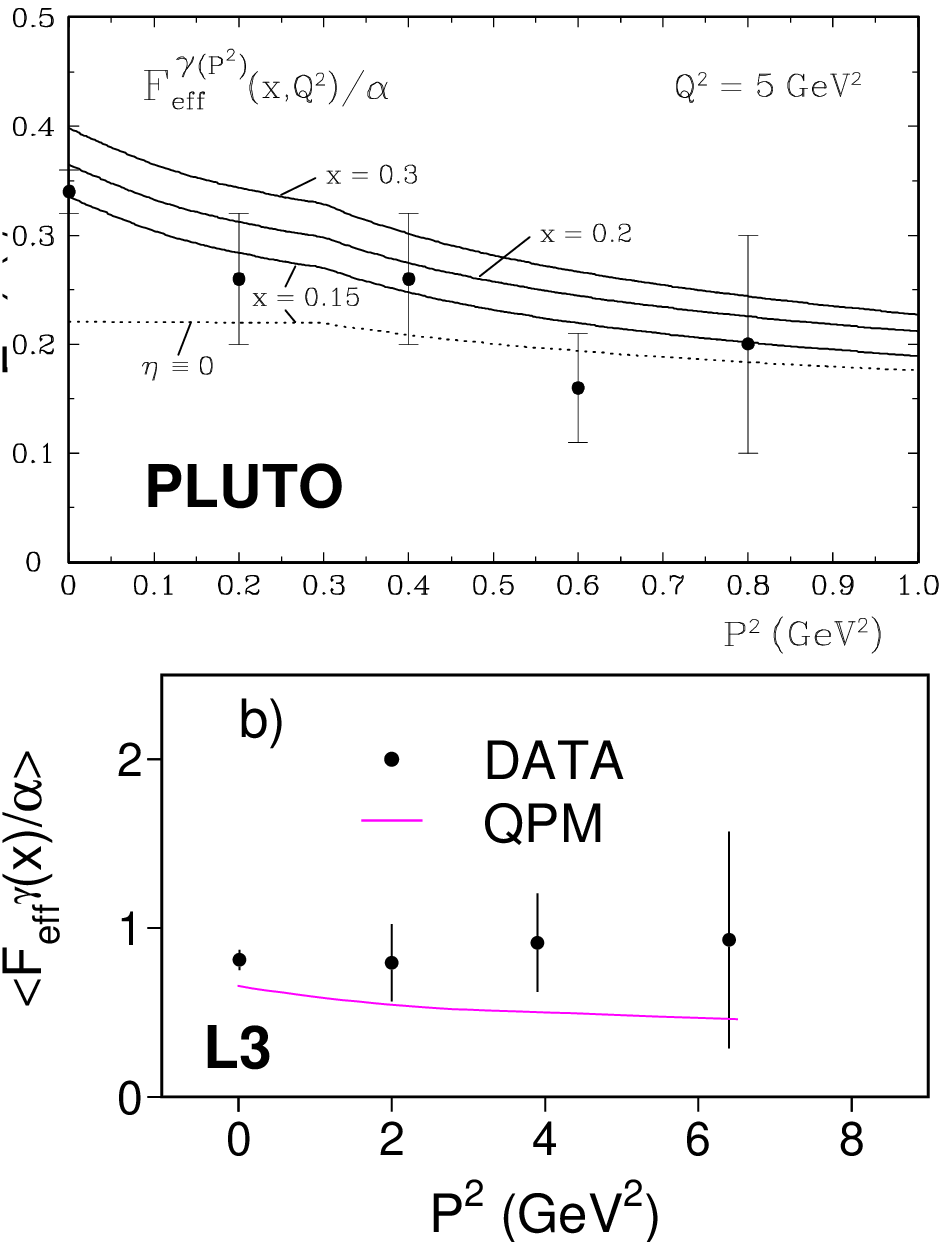,height=10cm}
\vspace*{-1.cm}
\caption{
The dependence of the effective photon structure function on the mass
$P^2$ of the probed photon.}
\label{fig:virtlep2}
\end{minipage}
\end{figure}
In the same experiment, the structure function of real photons was
also measured. Both results can be seen in figure~\ref{fig:virtlep1}
and within errors the structure function of the virtual photons is of
the same order of magnitude as that of the real ones. The effective
structure function is also presented as function of the virtuality of
the probed photon $P^2$ in figure~\ref{fig:virtlep2} and show very
little dependence on $P^2$ up to values of $\sim$ 6
GeV$^2$~\cite{pluto,virtl3}.

\subsection{What have we learned about the structure of the photon?}

Let us summarize what we have learned so far about the structure of
the photon. 
\begin{itemize}
\item At HERA one can see clear signals of the 2-component structure of 
quasi-real photons, a direct and a resolved part.
\item Virtual photons can also have a resolved part at low $x$ and fluctuate 
into $q\bar{q}$ pairs. 
\item Structure of virtual photons has been seen also at LEP.
\end{itemize}

\section{THE ANSWER}

Following the two sections on the structure of the proton and the
photon, let us remind ourselves again what our original question
was. At low $x$ we have seen that a $\gamma^*$ can have structure. Does
it still probe the proton in an $ep$ DIS experiment or does one of the
partons of the proton probe the structure of the $\gamma^*$?

The answer is just as Bjorken said: at low $x$ it does not
matter. Both interpretation are correct. The emphasis is however `at
low $x$'. At low $x$ the structure functions of the proton and of the
photon can be related through Gribov factorization~\cite{gribov}. By
measuring one, the other can be obtained from it through a simple
relation. This can be seen as follows.

Gribov showed~\cite{gribov} that the $\gamma\gamma$, $\gamma p$ and
$pp$ total cross sections can be related by Regge factorization as
follows:
\begin{equation}
\sigma_{\gamma\gamma}(W^2) 
= \frac{\sigma_{\gamma p}^2(W^2)}{\sigma_{pp}(W^2)}.
\end{equation}
This relation can be extended~\cite{who} to the case where one photon
is real and the other is virtual,
\begin{equation}
\sigma_{\gamma^*\gamma}(W^2,Q^2) 
= \frac{\sigma_{\gamma^* p}(W^2,Q^2)\sigma_{\gamma p}(W^2)}
{\sigma_{pp}(W^2)},
\end{equation}
or to the case where both photons are virtual,
\begin{equation}
\sigma_{\gamma^*\gamma^*}(W^2,Q^2,P^2) 
= \frac{\sigma_{\gamma^* p}(W^2,Q^2)\sigma_{\gamma^* p}(W^2,P^2)}
{\sigma_{pp}(W^2)}.
\end{equation}

\begin{figure}
\vspace*{-1cm}
\begin{minipage}{9.5cm}
\epsfig{file=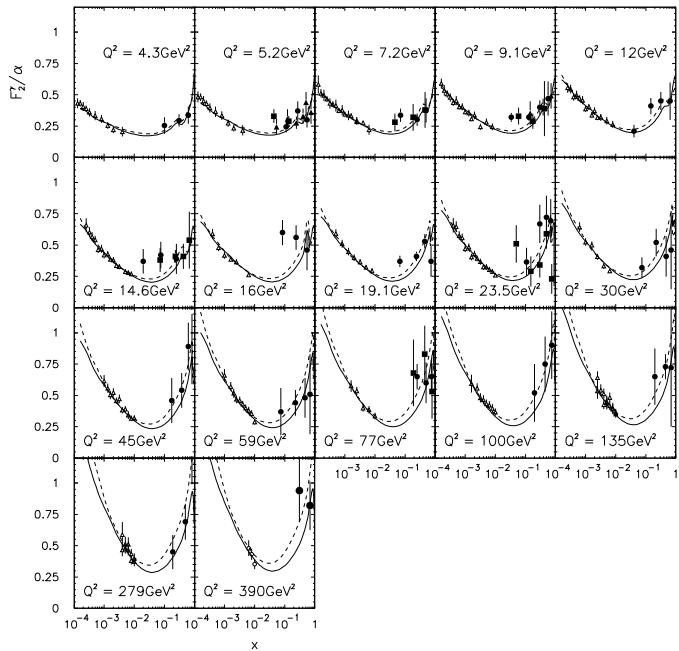,height=10cm}
\end{minipage}
\begin{minipage}{5.5cm}
\vspace*{-1.cm}
\caption{
The photon structure function, as function of $x$, for fixed $Q^2$
values as indicated in the figure. The full points are direct
measurements, and the open triangles are those obtained from $F_2^p$
through the Gribov factorization relation. The full line is the result
of a higher order fit and the dashed line is that of a leading order
parameterization.}
\label{fig:newf2}
\end{minipage}
\end{figure}

Since at low $x$ one has $ \sigma \approx \frac{4\pi^2\alpha}{Q^2} F_2
$, one gets the following relations between the proton structure
function $F_2^p$, the structure function of a real photon,
$F_2^\gamma$, and that of a virtual photon, $F_2^{\gamma^*}$:
\begin{equation}
F_2^\gamma(W^2,Q^2) = F_2^p(W^2,Q^2) \frac{\sigma_{\gamma p}(W^2)}
{\sigma_{pp}(W^2)},
\label{eq:fact}
\end{equation}
and
\begin{equation}
F_2^{\gamma^*}(W^2,Q^2,P^2) = \frac{4\pi^2\alpha}{P^2}\frac{
 F_2^p(W^2,Q^2) F_2^p(W^2,P^2)}
{\sigma_{pp}(W^2)}.
\end{equation}
The relation given in equation~(\ref{eq:fact}) has been
used~\cite{gal} to `produce' $F_2^\gamma$ `data' from well measured
$F_2^p$ data in the region of $x<$0.01, where the Gribov factorization
is expected to hold. The results are plotted in figure~\ref{fig:newf2}
together with direct measurements of $F_2^\gamma$. Since no direct
measurements exist in the very low $x$ region for $Q^2>4$ GeV$^2$, it
is difficult to test the relation. However both data sets have been
used for a global QCD leading order and higher order fits~\cite{galho}
to obtain parton distributions in the photon. Clearly there is a need
of more precise direct $F_2^\gamma$ data for such a study.

In any case, our answer to the question would be that at low $x$ the
virtual photon and the proton probe the structure of each other. In
fact, what one probes is the structure of the interaction. At high
$x$, the virtual photon can be assumed to be structureless and it
studies the structure of the proton.

\section{DISCUSSION - THE STRUCTURE OF THE INTERACTION}

We concluded in the last section that at low $x$ one studies the
structure of the interaction. Let us discuss this point more clearly.

We saw that in case of the proton at low $x$, the density of the
partons increases with decreasing $x$. Where are the partons located?
In the proton rest frame, Bjorken $x$ is directly related to the space
coordinate of the parton. The distance $l$ in the direction of the
exchanged photon is given by~\cite{ioffe},
\begin{equation}
l = \frac{1}{2m_px} \approx \frac{0.1{\rm fm}}{x}.
\end{equation}
Therefore partons with $x>$0.1 are in the interior of the proton,
while all partons with $x<$0.1 have no direct relation to the
structure of the proton. The low $x$ partons describe the properties
of the $\gamma^* p$ interaction.

\begin{figure}[htb]
\begin{minipage}{7.5cm}
\epsfig{file=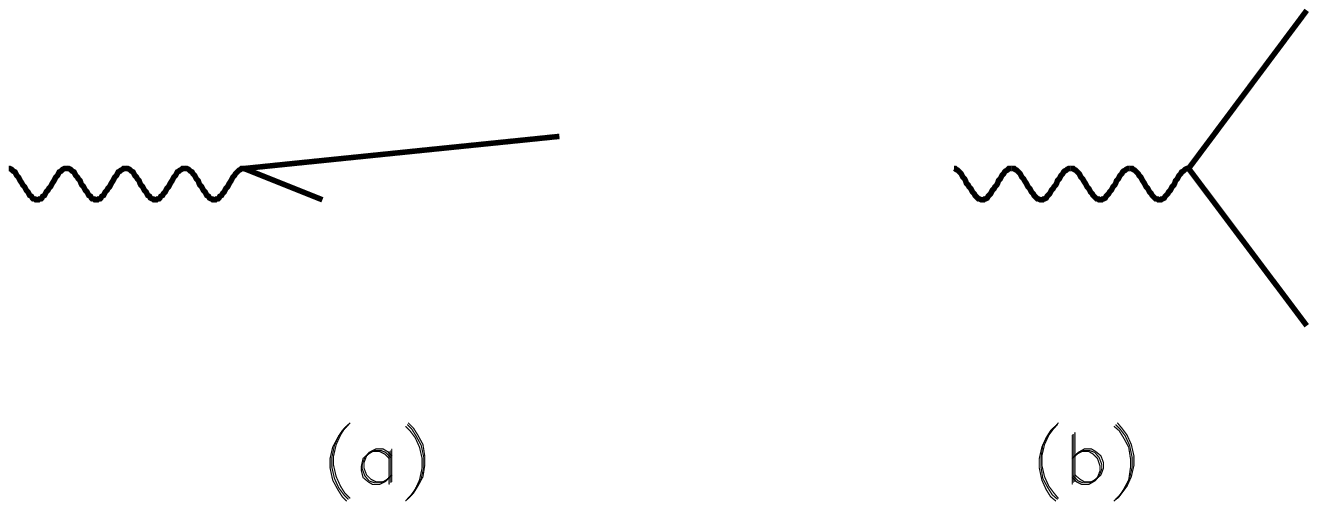,height=3cm}
\vspace*{-1.cm}
\caption{
Fluctuation of the photon into a $q\bar{q}$ pair in (a) an asymmetric
small $k_T$ configuration, (b) a symmetric large $k_T$
configuration.}
\label{fig:ajm}
\end{minipage}
\hspace*{4mm}
\begin{minipage}{7.5cm}
\epsfig{file=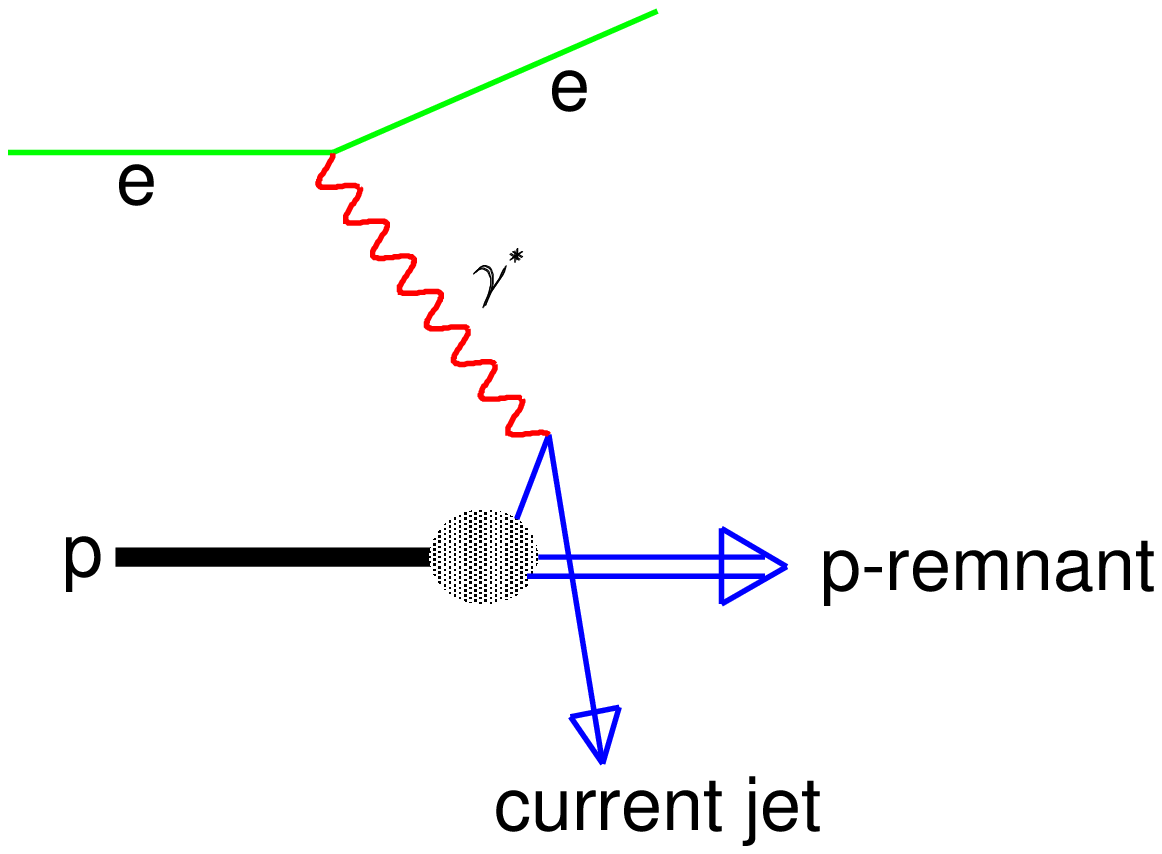,height=5cm}
\vspace*{-1.cm}
\caption{A DIS diagram, as seen from the point of view of a photon 
fluctuating into an asymmetric pair, in which the fast quark becomes
the current jet.}
\label{fig:dis-ajm}
\end{minipage}
\end{figure}

How can we describe a $\gamma^* p$ interaction at low $x$? It occurs
in two steps: first the virtual photon fluctuates into a $q\bar{q}$
pair and then the configuration of this pair determines if the
interaction is `soft' or `hard'~\cite{ajm}. The soft process is the
result of a large spatial configuration in which the photon fluctuates
into an asymmetric small $k_T$ $q\bar{q}$ pair. The hard nature of the
interaction is obtain when the fluctuation is into a small
configuration of a symmetric $q\bar{q}$ pair with large $k_T$. The two
configurations are shown in figure~\ref{fig:ajm}.
At $Q^2$=0 the asymmetric configuration is dominant and the large
color forces produce a hadronic component which interacts with the
proton and leads to hadronic non-perturbative soft physics. The
symmetric component contributes very little; the high $k_T$
configuration is screened by color transparency (CT). At higher $Q^2$
the contribution of the symmetric small configuration gets
bigger. Each one still contributes little because of CT, but the phase
space for such configurations increases. Nevertheless, the asymmetric
large configuration is also still contributing and thus both soft and
hard components are present. Another way to see this interplay is by
looking at the diagram in figure~\ref{fig:dis-ajm}. In a simple QPM
picture of DIS, the fast quark from the asymmetric configuration
becomes the current jet while the slow quark interacts with the proton
in a soft process. Thus the DIS process looks in the $\gamma^* p$
frame just like the $Q^2$=0 case. This brings the interplay of soft
and hard processes.

\section{CONCLUSION}
\begin{itemize}
\item
In DIS experiments at low $x$ one studies the `structure' of the
$\gamma^* p$ interaction.
\item
In order to study the interior structure of the proton, one needs to
measure the high $x$ high $Q^2$ region. This will be done at HERA after the 
high luminosity upgrade.
\end{itemize}

\vspace{1cm}
\noindent
I would like to thank Professors K. Maruyama and H. Okuno for
organizing a pleasant and lively Symposium. Special thanks are due to
Professor K. Tokushuku and his group for being wonderful hosts during
my visit in Japan. Finally I would like to thank Professor
H. Abramowicz for helpful discussions.

\end{document}